\documentclass[traditabstract]{aa}
\usepackage[dvips]{graphicx}
\usepackage{txfonts}
\usepackage{natbib}
\bibpunct{(}{)}{;}{a}{}{,}
\begin{document}

\title{An unbiased measurement of the UV background and its evolution via the
  proximity effect in quasar spectra
\thanks{Based on data collected at the European Southern Observatory and obtained from the ESO Science Archive.}}

\titlerunning{An unbiased measurement of the UV background}
\author{Aldo Dall'Aglio \and Lutz Wisotzki \and G\'{a}bor Worseck}

\institute{Astrophysikalisches Institut Potsdam, An der Sternwarte 16, D-14482 Potsdam, Germany\\
 \email{adaglio@aip.de}}

\date{Received ... /  Accepted ...}

\abstract{We investigated a set of high-resolution ($R\sim45\ 000$), high
signal-to-noise ($S/N\sim70$) quasar spectra to search for the signature of
the so-called proximity effect in the \ion{H}{i} Ly$\alpha$ forest.  The
sample consists of 40 bright quasars with redshifts in the range $2.1 < z <
4.7$. Using the flux transmission statistic, we
determined the redshift evolution of the \ion{H}{i} effective optical depth in
the Lyman forest between $2\la z\la 4.5$, finding good agreement with previous
measurements based on smaller samples. We also see the previously reported dip
in $\tau_\mathrm{eff}(z)$ around redshift $z\sim 3.3$, but as the significance
of that feature is only 2.6$\sigma$, we consider this detection
tentative. Comparing the flux transmission near each quasar with what was expected
from the overall trend of $\tau_\mathrm{eff}(z)$, we clearly detect the
proximity effect not only in the combined quasar sample, but also towards each
individual line of sight at high significance, albeit with varying strength.
We quantify this strength using a simple prescription based on a fiducial
value for the intensity of the metagalactic UV background (UVB) radiation
field at 1~Ryd, multiplied by a free parameter that varies from QSO to QSO.
The observed proximity effect strength distribution (PESD) is asymmetric, with
an extended tail towards values corresponding to a weak effect. We demonstrate
that this is not simply an effect of gravitational clustering around
quasars, as the same asymmetry is already present in the PESD predicted for
purely Poissonian variance in the absorption lines. We present the results of
running the same analysis on simulated quasar spectra generated by a simple
Monte-Carlo code. Comparing the simulated PESD with observations, we argue
that the standard method of determining the UVB intensity $J_{\nu_0}$ by
averaging over several lines of sight is heavily biased towards high values of
$J_{\nu_0}$ because of the PESD asymmetry. 
Using instead the \emph{mode} of the PESD provides an estimate of
$J_{\nu_0}$ that is unbiased with respect to his effect.  For our
sample we get a modal value for the UVB intensity of $\log J_{\nu_0} = -21.51\pm 0.15$ (in units
of
$\mathrm{erg}\,\mathrm{cm}^{-2}\,\mathrm{s}^{-1}\,\mathrm{Hz}^{-1}\,\mathrm{sr}^{-1}$)
for a median quasar redshift of 2.73. With $J_{\nu_0}$ fixed we then corrected
$\tau_\mathrm{eff}$ near each quasar for local ionisation and estimated the
amount of excess \ion{H}{i} absorption attributed to gravitational
clustering. On scales of $\sim 3$~Mpc, only a small minority of quasars show
substantial overdensities of up to a factor of a few in $\tau_\mathrm{eff}$;
these are exactly the objects with the weakest proximity effect signatures.
After removing those quasars residing in overdense regions, we redetermined
the UVB intensity using a hybrid approach of sample averaging and statistical
correction for the PESD asymmetry bias, arriving at $\log J_{\nu_0} = -21.46^{+0.14}_{-0.21}$. 
This is the
most accurate measurement of $J_{\nu_0}$ to date. We present a new diagnostic
based on the shape and width of the PESD that strongly supports our
conclusion that there is no systematic overdensity bias for the proximity
effect. This additional diagnostic breaks the otherwise unavoidable degeneracy
of the proximity effect between UVB and overdensity.  We then applied our
hybrid approach to estimate the redshift evolution of the UVB intensity and
found tentative evidence of a mild decrease in $\log J_{\nu_0}$ with
increasing redshift, by a factor of $\sim 0.4$ from $z=2$ to $z=4$. Our
results are in excellent agreement with earlier predictions for the evolving UVB
intensity, and they also agree well with other methods of
estimating the UVB intensity.  In particular, our measured UVB evolution is much
slower than the change in quasar space densities between $z=4$ and $z=2$,
supporting the notion of a substantial contribution of star-forming galaxies
to the UVB at high redshift. }

\keywords{diffuse radiation -- intergalactic medium -- quasars: absorptionlines}
\maketitle

\begin{figure*}
\sidecaption
\includegraphics[width=12cm]{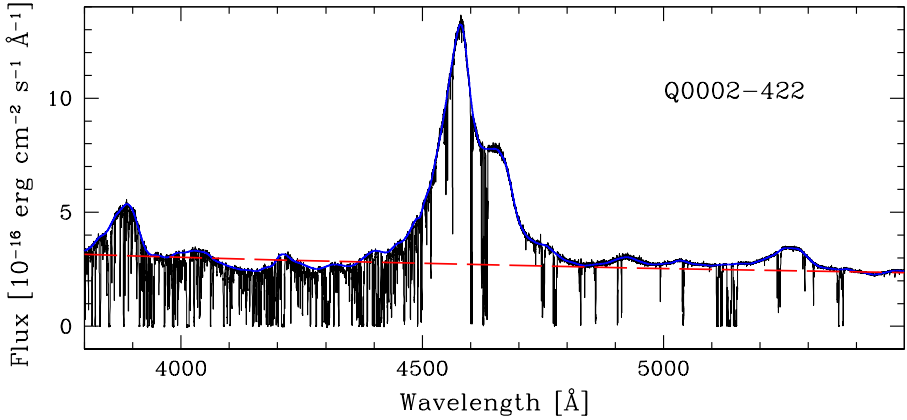}
\caption{A typical quasar spectrum from our sample (Q~0002$-$422 $z=2.768$) together with 
the power law continuum fit (dashed red line) and the local continuum fit (blue line).}
\label{fig:qso}
\end{figure*}

\section{Introduction}

High-resolution spectra of high-redshift, quasi-stellar objects (QSOs) show a
plethora of narrow absorption lines from various species along their lines of
sight through the intergalactic medium (IGM). Most absorption lines blueward
of an Ly$\alpha$ emission line stem from the Ly$\alpha$ transition of
intervening neutral hydrogen, giving rise to the Ly$\alpha$ forest
\citep{sargent80,weymann81,rauch98}. High-resolution spectra of the Ly$\alpha$
forest enabled a statistical analysis of the absorbers, as well as their
physical properties \citep[e.g.][]{kim01,schaye03,kim04}.

The vast majority of \ion{H}{i} Ly$\alpha$ absorbers are optically thin to
ionising photons and are thus kept highly photoionised by the metagalactic UV
background (UVB) generated by the overall population of quasars and
star-forming galaxies \citep{haardt96,fardal98,haardt01}. The UVB intensity 
at a given frequency is determined by the evolving population of UV sources,
modified by cosmological expansion and absorption
\citep[e.g.][]{dave99}. It is important to determine the amplitude of the UVB
and its spectral shape as a function of redshift in order to constrain the
relative contributions of quasars and star-forming galaxies to the UVB, as
well as to provide a key ingredient to numerical simulations of structure
formation.

There are different ways to constrain the \ion{H}{i} UVB photoionisation rate
in the IGM. Assuming its spectral energy distribution it is possible to infer
the UVB intensity at the Lyman limit $J_{\nu_0}$.  First, the UVB can be
predicted by integrating the contributions of the observed source
population. Given the luminosity functions of the sources, a characteristic intrinsic
spectral energy distribution and the observed absorber distribution functions
in the IGM, the amplitude and spectral shape of the UVB can be calculated
numerically as the radiation is filtered by the IGM \citep{bechtold87}.  In
particular, comparing these models to observations provides constraints on the
relative contributions of quasars \citep{haardt96,fardal98} and star-forming
galaxies \citep{haardt01}.

Furthermore, the \ion{H}{i} photoionisation rate at $2<z<4$ has been
constrained by matching the Ly$\alpha$ forest absorption and the IGM
temperature evolution from numerical simulations of structure formation to
observations
\citep{rauch97,theuns98,mcdonald00,meiksin04,tytler04,bolton05,kirkman05,jena05}.

A more direct approach to estimate the UVB photoionisation rate is based on
the so-called proximity effect. In the vicinity of a luminous quasar the UV
radiation field is expected to be enhanced, resulting in a statistical deficit
of Ly$\alpha$ absorbers near the quasar compared to far away from it along the line
of sight \citep{weymann81,carswell82,murdoch86}. Knowing the UV luminosity of
the quasar, the deficit of lines (or generally Ly$\alpha$ opacity) yields an
estimate of the UVB intensity at the \ion{H}{i} Lyman limit
\citep{carswell87,bajtlik88}. Traditionally, the proximity effect has been
considered as a statistical effect and large samples of up to $\sim 100$
quasars have been compiled to measure the mean UVB at $2\la z\la 4$
\citep[e.g.][]{bajtlik88,lu91,giallongo96,cooke97,scott00,liske01}.
Nevertheless, the proximity effect can also be detected towards individual
quasars \citep[e.g.][]{williger94,cristiani95,lu96,savaglio97,adaglio08}.

Proximity effect analyses at different redshifts can constrain the redshift
evolution of the UVB, showing that it broadly peaks at $z\sim 3$. The decrease
in the UVB at $z< 2$ \citep{kulkarni93,scott02} and $z\ga 4$
\citep{williger94,lu96,savaglio97} is likely due to the declining space
density of quasars \citep{haardt96,fardal98}. Whereas the UVB evolution  at low redshifts is
sufficiently fast to be detectable within a single sample
\citep{scott02}, there is presently little direct evidence of evolution 
within the redshift range $2\la z\la 4$.

It is currently an open issue whether the measured UVB at $z\la 3$ is
consistent with the integrated emission of quasars, or whether a significant
share of ionising photons has to be provided by star-forming galaxies
\citep[e.g.][]{madau99,sokasian03,schirber03,hunt04}.  One difficulty lies in
the fact that the UVB estimates via the proximity effect are subject to
several systematic uncertainties, all of which can lead to an overestimate of the
inferred UV background:
\begin{enumerate}
\item \textit{Quasar variability:} The size of the proximity effect zone
scales as the average quasar luminosity over the photoionisation equilibrium timescale in
the IGM ($\sim 10^4$~yr). Quasar variability on shorter timescales will tend
to overestimate the UVB due to Malmquist bias in the selected quasar samples
\citep{schirber04}. It is not known how important this effect is in practice.
\item \textit{Underestimated quasar redshifts:} Quasar emission redshifts
determined from broad high-ionisation lines are likely underestimated
\citep[e.g.][]{gaskell82,richards02}, causing the UVB to be overestimated by a
factor $\sim 3$ \citep{espey93}. However, this effect can be largely avoided
by using low-ionisation lines to determine systemic redshifts.
\item \textit{Overdense environment:} In the standard ionisation model of the
proximity effect \citep{bajtlik88}, the Ly$\alpha$ forest line density is
extrapolated into the quasar vicinity, assuming that the matter distribution
is not altered by the presence of the QSO.  If quasars typically reside in
high intrinsic overdensities instead, the UVB will be overestimated by a
factor of up to $\sim 3$ \citep{loeb95}. Conversely, the disagreement between
the UVB estimates from the proximity effect and those obtained by matching simulations to the
mean Ly$\alpha$ forest absorption has recently been used as an argument that quasars
are surrounded by substantial overdensities \citep{rollinde05,guimares07,giguere08}.
\end{enumerate}

In the present paper we pursue a new approach to employ the proximity effect
as a tool to investigate the UVB, based on measuring its
signature towards individual QSO sight lines. We demonstrated recently
\citep[][ hereafter Paper~I]{adaglio08} that indeed essentially all quasars
show this signature, even at relatively low spectral resolution. Here we
follow up on this finding, now based exclusively on high resolution, high S/N
quasar spectra. The plan of the paper is as follows.  We begin with a
brief description of our spectroscopic data (Sect.~\ref{txt:data}). In
Sect.~\ref{txt:sims} we explain the Monte Carlo simulations used to assess
uncertainties and to interpret the observed proximity effect strength
distribution. We then determine the redshift
evolution of the effective optical depth in the Ly$\alpha$ forest in Sect.~\ref{txt:tau_evol}, followed by
a comparison between the two most commonly adopted methods of revealing the
proximity effect in Sect.~\ref{txt:method}. We very briefly report on the
analysis of the combined QSO sample in Sect.~\ref{txt:comb_PE}. In
Sect.~\ref{txt:sglos} we investigate the proximity effect in individual
sightlines, showing that the traditional sample-combining method is biased.
We then estimate the excess Ly$\alpha$ absorption near quasars 
(Sect.~\ref{txt:data_overd}), and use this to constrain
the redshift evolution of the UVB (Sect.~\ref{txt:uvb}). We present our
conclusions in Sect.~\ref{txt:concl}.

Throughout this paper we assume a flat universe with Hubble constant
$H_0=70$~$\mathrm{km}\,\mathrm{s}^{-1}\,\mathrm{Mpc}^{-1}$ and density
parameters
$\left(\Omega_\mathrm{m},\Omega_\Lambda\right)=\left(0.3,0.7\right)$. All
distances are expressed in physical units.

\section{Data}\label{txt:data}

\subsection{Sample description and data reduction}\label{txt:desc}

The sample consists of 40 quasar sight lines observed with the UV-Visual Echelle
Spectrograph \citep[UVES,][]{dekker00} at VLT/UT2 on Cerro Paranal, Chile. All
40 spectra were taken from the ESO archive and are publicly available. We
selected the QSO sample based on the following criteria: (i) a minimum of 10
exposures and (ii) a complete or nearly complete coverage of the Ly$\alpha$
forest region especially close to the quasar. Table~\ref{tab:data} lists the
complete sample of quasars with adopted redshifts (Sect.~\ref{txt:sys_z}) and
quasar luminosities at the Lyman limit (Sect.~\ref{txt:cont}).

The data were reduced within the ECHELLE/UVES environment (version 2.2) of the
software package MIDAS and following the procedures described in
\citet{kim04}.  After the bias and inter-order background subtraction from
each science frame, an optimal extraction of the spectrum was performed, order
by order, assuming a Gaussian distribution along the spatial direction. The sky
was estimated and subtracted together with the Gaussian fit, maximising the quality of the
extracted spectrum \citep[e.g.][]{kim01}. Cosmic rays were identified and
removed using a median filter. The wavelength calibration was performed
with the available ThAr lamp frames. Relative flux calibration was performed
using the master-response functions publicly available on the ESO website
 and accounting for different airmasses.
This allowed us to place our spectra onto a relative
flux scale reasonably well, while the absolute scale had to be tied to external photometry
(Sect.~\ref{txt:mag}). In each final coadded spectrum the individual
extractions were weighed corresponding to their S/N and re-sampled on
0.05~\AA\ bins.  The resolving power is $\sim 45\,000$ in the regions of
interest, corresponding to a velocity resolution of $\sim
6$~$\mathrm{km}\,\mathrm{s}^{-1}$ (FWHM). The wavelengths in the final spectra
are vacuum heliocentric, and the fluxes are corrected for galactic reddening.
An example of a final spectrum as well as our continuum definitions
(Sect.~\ref{txt:cont}) is shown in Fig.~\ref{fig:qso}.

\begin{table}
\tiny\centering
\caption[]{List of analysed QSOs, ordered by redshift.}
\label{tab:data}
\begin{tabular}{lcccccl}
\hline\hline\noalign{\smallskip}
QSO &   Mag & Filter & S/N & $f_{\nu_0}^{\ddag}$&$\log L_{\nu_0}^{\star}$& ${z_\mathrm{em}}^\dagger$\\
\noalign{\smallskip}\hline\noalign{\smallskip}
\object{HE~1341$-$1020}             & 17.1 & $B$ & $70$  & 139 & 31.17  & 2.137 \\
\object{Q~0122$-$380}               & 16.7 & $B$ & $60$  & 291 & 31.51  & 2.192 \\
\object{PKS~1448$-$232}             & 17.0 & $V$ & $70$  & 282 & 31.50  & 2.222 \\
\object{PKS~0237$-$233}             & 16.6 & $V$ & $100$ & 750 & 31.93  & 2.224 \\
\object{HE~0001$-$2340}             & 16.7 & $V$ & $60$  & 403 & 31.68  & 2.278 \\
\object{HE~1122$-$1648}$\ ^{\rm{a}}$& 16.5 & $V$ & $120$ & 866 & 32.05  & 2.407 \\
\object{Q~0109$-$3518}              & 16.4 & $B$ & $70$  & 404 & 31.72  & 2.406 \\
\object{HE~2217$-$2818}             & 16.0 & $V$ & $80$  & 796 & 32.02  & 2.414 \\
\object{Q~0329$-$385}               & 17.0 & $V$ & $50$  & 226 & 31.48  & 2.437$\ ^{\rm{b}}$ \\
\object{HE~1158$-$1843}             & 16.9 & $V$ & $60$  & 268 & 31.56  & 2.459 \\
\object{Q~2206$-$1958}              & 17.3 & $V$ & $70$  & 255 & 31.57  & 2.567 \\
\object{Q~1232$+$0815}              & 18.4 & $r$ & $50$  & 055 & 30.91  & 2.575 \\
\object{HE~1347$-$2457}             & 16.8 & $B$ & $60$  & 772 & 32.06  & 2.615 \\
\object{HS~1140$+$2711}$\ ^{\rm{a}}$& 16.7 & $r$ & $80$  & 581 & 31.94  & 2.628 \\
\object{Q~0453$-$423}               & 17.1 & $V$ & $70$  & 168 & 31.41  & 2.664 \\
\object{PKS~0329$-$255}             & 17.1 & $V$ & $50$  & 243 & 31.58  & 2.706 \\
\object{Q~1151$+$0651}              & 18.1 & $r$ & $50$  & 100 & 31.21  & 2.758 \\
\object{Q~0002$-$422}               & 17.2 & $V$ & $60$  & 346 & 31.76  & 2.769 \\
\object{Q~0913$+$0715}              & 17.8 & $r$ & $60$  & 181 & 31.48  & 2.788 \\
\object{HE~0151$-$4326}             & 17.2 & $B$ & $70$  & 499 & 31.92  & 2.787 \\
\object{Q~1409$+$095}               & 18.6 & $r$ & $40$  & 067 & 31.06  & 2.843 \\
\object{HE~2347$-$4342}             & 16.7 & $V$ & $100$ & 630 & 32.05  & 2.886 \\
\object{Q~1223$+$1753}              & 18.1 & $r$ & $40$  & 794 & 31.16  & 2.955 \\
\object{Q~0216$+$080}$\ ^{\rm{a}}$  & 18.1 & $V$ & $40$  & 084 & 31.20  & 2.996 \\
\object{HE~2243$-$6031}             & 16.4 & $V$ & $100$ & 776 & 32.17  & 3.012 \\
\object{CTQ~0247}                   & 17.4 & $V$ & $70$  & 239 & 31.66  & 3.025 \\
\object{HE~0940$-$1050}             & 16.4 & $V$ & $50$  & 691 & 32.13  & 3.089 \\
\object{Q~0420$-$388}               & 16.9 & $V$ & $100$ & 426 & 31.93  & 3.120 \\
\object{CTQ~0460}                   & 17.5 & $V$ & $60$  & 281 & 31.76  & 3.141 \\
\object{Q~2139$-$4434}              & 17.7 & $V$ & $40$  & 117 & 31.39  & 3.207 \\
\object{Q~0347$-$3819}              & 17.7 & $V$ & $60$  & 213 & 31.66  & 3.229 \\
\object{PKS~2126$-$158}             & 17.0 & $V$ & $60$  & 380 & 31.92  & 3.285$\ ^{\rm{c}}$ \\
\object{Q~1209$+$0919}              & 18.6 & $r$ & $40$  & 120 & 31.42  & 3.291$\ ^{\rm{c}}$ \\
\object{Q~0055$-$2659}              & 17.5 & $V$ & $60$  & 245 & 31.81  & 3.665$\ ^{\rm{c}}$ \\
\object{Q~1249$-$0159}$\ ^{\rm{a}}$ & 18.6 & $g$ & $60$  & 071 & 31.27  & 3.668 \\
\object{Q~1621$-$0042}              & 17.3 & $r$ & $70$  & 307 & 31.91  & 3.709 \\
\object{Q~1317$-$0507}              & 17.7 & $R$ & $50$  & 084 & 31.35  & 3.719 \\
\object{PKS~2000$-$330}$\ ^{\rm{a}}$& 17.0 & $V$ & $60$  & 178 & 31.69  & 3.786 \\
\object{BR~1202$-$0725}             & 17.8 & $R$ & $100$ & 460 & 32.24  & 4.697 \\
\object{Q~1451$-$1512}             & 17.3 & $I$ & $60$  & 332 & 32.12  & 4.766 \\
\noalign{\smallskip}\hline\noalign{\smallskip}
\end{tabular}
\begin{list}{}{}
\item[$\ddag$:] Lyman limit flux in units of $\mu\mathrm{Jy}$ with uncertainties around 7\%.
\item[$\star$:] Lyman limit luminosities in units of $\mathrm{erg}\,\mathrm{s}^{-1}\,\mathrm{Hz}^{-1}$ 
with uncertainties of the order of $\sigma_{\log L_{\nu_0}}\simeq^{+0.05}_{-0.1}$ dex.
\item[$\dagger$:] Systemic redshift estimated from \ion{Si}{ii}$+$\ion{O}{i} at 
$\lambda_\mathrm{rest}=1305.77$\AA\ \citep{morton03}. Redshift uncertainty is $\sigma_z=0.003$.
\item[a:] Gaps in the spectral range of the Ly$\alpha$ forest.
\item[b:] Redshift measured from \ion{C}{iv} emission line and corrected according to \citet{richards02}.
\item[c:] Redshift taken from Paper~I.
\end{list}
\end{table}

\subsection{Quasar magnitudes}\label{txt:mag}

Table~\ref{tab:data} provides our adopted apparent magnitudes and Lyman limit
fluxes. The photometric data were collected from various sources, in
particular SDSS, \citet{veron06}, \citet{rollinde05}, \citet{cooke97},
\citet{ellison05}, and Paper~I. Where more than one measurement was available,
the data were consistent to within $\sigma_m < 0.1$ mag. Note that while
knowledge of the Lyman limit luminosity is indispensable for the quantitative
interpretation of the proximity effect, even perfectly simultaneous high-S/N
photometry would not account for the effects of quasar variability averaged
over the photoionisation time scale. By randomly changing the flux scale of
some objects by up to 20\% we confirmed that our conclusions concerning the
proximity effect are not affected by photometric uncertainties.

\subsection{Continuum definition}\label{txt:cont}

In order to search for the proximity effect, we needed to convert the quasar
spectra into continuum-rectified transmission spectra. We also needed to know
the quasar flux at the Lyman limit. To achieve this we determined two types of
continua: (i) a global power law ($f_\nu \propto \nu^\alpha$), excluding
emission and absorption regions, used to estimate the quasar flux at the Lyman
limit and (ii) a more local estimate that also includes the broad emission
lines as a quasi-continuum, used to compute the transmission spectrum. We
constrained the Lyman limit flux via a power law continuum fit since in most
cases this wavelength is either not covered by the observations or is
located in the far blue range of the spectrum where the S/N significantly
drops.

We developed an algorithm to automatically fit the local continuum, building
on the work by \citet{young79} and \citet{carswell82}. A cubic spline was
interpolated on adaptive intervals along the spectrum.  The fixpoints for the
spline interpolation were chosen starting from a regular sampling of the
spectrum with a binning that becomes finer whenever the slope of the computed
continuum exceeds a given threshold. This was done in order to better
reproduce the wings of emission lines.

\begin{figure}
\resizebox{\hsize}{!}{\includegraphics*{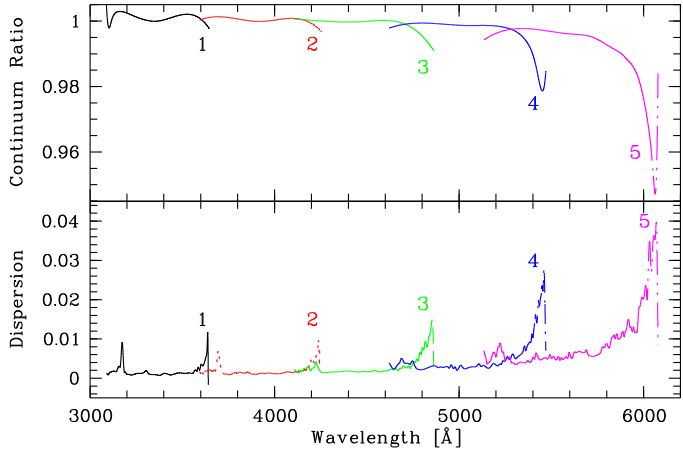}}
\caption{Top panel: Average ratio between the fitted and input continuum for
the five sets of 500 simulated QSOs at $z$ = 2.0, 2.5, 3.0, 3.5, 4.0 with $S/N
= 70$, denoted with numbers 1, 2, 3, 4 and 5 respectively. Bottom panel:
Standard deviation profiles relative to the above systematic bias.}
\label{fig:contchange}
\end{figure}

To assess the continuum uncertainties we proceeded in the same way as
described in Paper~I. With the help of Monte Carlo simulations we computed the
average ratio between the recovered and the input continua at different
redshifts. The results are presented in Fig.~\ref{fig:contchange}. The top
panel shows the mean systematic error of the fitted continuum with respect to
the input continuum for five different redshifts. The bottom panel presents
the intrinsic uncertainties of the continuum.  Due to the high resolution and
the high S/N of the spectra, the statistical error associated with continuum
placement is limited to a few percent at most in the Ly$\alpha$ line wing. The
automatically fitted continuum was corrected for the systematic bias and used
to compute the transmission $T=F_{\rm{qso}}/F_{\rm{cont}}$.

Regarding the Lyman limit fluxes, the uncertainties are of the order of $7\%$, 
dominated by the somewhat subjective choice of the continuum wavelength ranges
redward of the Ly$\alpha$ emission used to fit the power laws.

\subsection{Systemic quasar redshift}\label{txt:sys_z}

The spectral range covered by the UVES observations can maximally extend from
$\sim$3\,000 to $\sim$10\,000 \AA. However, not all QSOs have such wide
wavelength coverage. Most of them extend far enough to the red of Ly$\alpha$ to allow
the detection of several emission lines (\ion{Si}{ii}+\ion{O}{i}, \ion{C}{ii},
\ion{Si}{iv}+\ion{O}{iv]}, \ion{C}{iv}) from which a redshift can be
measured. Several authors \citep[e.g.][]{gaskell82, tytler92,richards02}
pointed out the existence of a systematic shift between high and low
ionisation lines attributed to the relative motion of gas near the AGN. In
order to measure a redshift as close as possible to the systemic one, we used
low-ionisation lines if these were available.  In almost all cases we adopted
\ion{Si}{ii}+\ion{O}{i} measurements for the final emission redshift, as
listed in Table~\ref{tab:data}. For four objects (PKS~2126$-$158,
Q~0329$-$385, Q~1209$+$0919 and Q~0055$-$2659) this was not possible due to
either a non-detection of these lines or lack of wavelength coverage. For
PKS~2126$-$158, Q~1209$+$0919 and Q~0055$-$2659 we adopted the redshifts
based on \ion{Si}{ii}+\ion{O}{i} from the low-resolution spectra in Paper~I,
while for Q~0329$-$385 we could only take the \ion{C}{iv} based redshift and
corrected it for the mean systematic shift determined by \citet{richards02}.

\begin{figure*}
\sidecaption
\includegraphics[width=12cm]{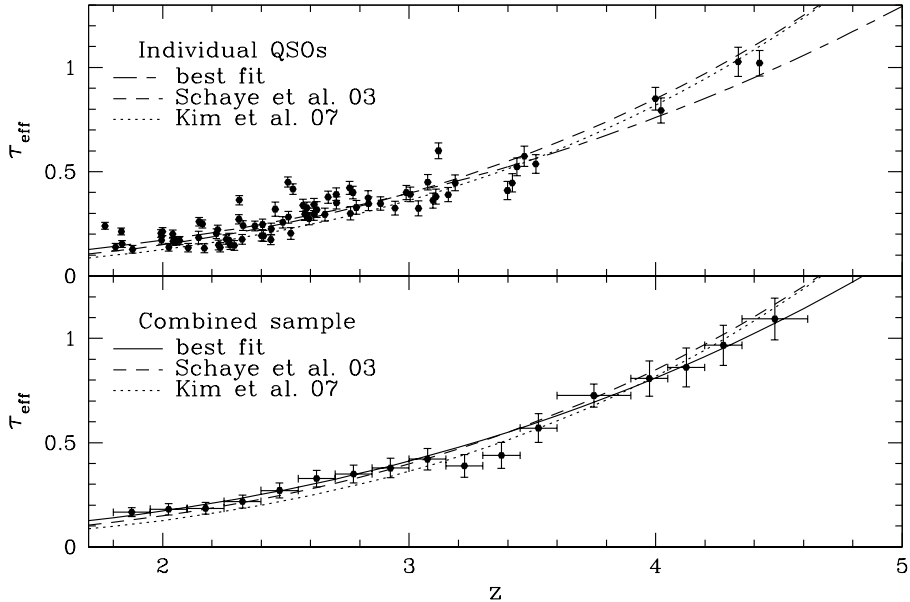}
\caption{Observed effective optical depth as a function of redshift after
removal of pixels contaminated by Ly$\alpha$ absorption lines with damping
wings, Ly limit systems and the proximity effect. Measurements from each
individual quasar are shown in the upper panel, while those from the combined
sample of QSOs are shown in the lower panel. The solid curve shows the power
law least-squares fit to the data while the dashed curve indicates the
evolution of the Ly$\alpha$ optical depth fitted by \citet{schaye03} and the
dotted line that by \citet{kim07}.}
\label{fig:tauevol}
\end{figure*}

\section{Monte Carlo simulations of artificial spectra}\label{txt:sims}

A deeper understanding of statistical and systematic effects contributing to
the proximity effect signature is made possible by comparing observations with
simulated spectra. Typically, numerical 3D simulations are successful in
reproducing the properties of the Ly$\alpha$ forest at any redshift.  However,
such computations are very time consuming and their major shortcoming is the
limited redshift coverage of the simulated sight lines, especially at high
resolution.

We developed a straightforward Monte Carlo code to generate synthetic spectra
 and employed the simulations for three main purposes. First, to
estimate the influence of the quasar continuum placement in the spectra
(Sect.~\ref{txt:cont}); then to compute the line-of-sight variations in the
evolution of the optical depth in the Ly$\alpha$ forest
(Sect.~\ref{txt:tau_evol}); and finally to study the influence of 
the fact that only a finite number of absorbers contributes to the
measured optical depth (Sect.~\ref{txt:errors}).

The procedure used to generate synthetic spectra is based on the assumption
that the Ly$\alpha$ forest is well represented by three observed
distributions:
\begin{enumerate}
\item{The line number density distribution, approximated by a power law of the form 
$\mathrm{d}n / \mathrm{d}z \propto (1+z)^{\gamma}$ where $\gamma$ will be measured in 
Sect.~\ref{txt:tau_evol}.}
\item{The column density distribution, given by 
$f(N_{\mbox{\scriptsize\ion{H}{i}}}) \propto N_{\mbox{\scriptsize\ion{H}{i}}}^{-\beta}$ 
where the slope is $\beta \simeq 1.5$ \citep{kim01}.}
\item{The Doppler parameter distribution, given by 
 $\mathrm{d}n / \mathrm{d}b \propto b^{-5}\mathrm{exp}\left[{-{b_{\sigma}^4}/{b^4}}\right]$ 
where  $b_{\sigma}\simeq 24\;\mathrm{km/s}$ \citep{kim01}.}
\end{enumerate}
The simulated absorbers have column densities within the range
$10^{12}<N_{\mbox{\scriptsize\ion{H}{i}}}<10^{18}$ cm$^{-2}$ and Doppler
parameters between $10<b<100$ km/s. The slope of the column density
distribution was fixed to $\beta = 1.5$.

For our purposes these approximations are sufficient to yield the accuracy we
need in the following analysis. Every simulated sight line was populated with
absorption features drawn from the above distributions until the computed
effective optical depth was consistent with its evolution presented in
Sect.~\ref{txt:tau_evol}. Once the simulated transmission was computed, an
artificial quasar spectral energy distribution, including emission lines of
varying strengths and widths, was generated via the principal components method
described by \citet{suzuki06} and already employed in Paper~I. Gaussian noise was added to the
final quasar spectrum, in order to match the S/N level of our observed
objects.

\begin{table}
\caption{Effective optical depth in the Ly$\alpha$ forest of our combined QSO sample.}
\label{tab:tau_comb}
\centering
\begin{tabular}{cccc}
\hline\hline\noalign{\smallskip}
$<z>$& $\Delta z$ & $\tau_\mathrm{eff}$ & $\sigma_{\tau_\mathrm{eff}}$ \\
\noalign{\smallskip}\hline\noalign{\smallskip}
 1.874 &  0.15 &  0.166 &   0.021\\
 2.024 &  0.15 &  0.180 &   0.026\\
 2.174 &  0.15 &  0.185 &   0.028\\
 2.324 &  0.15 &  0.217 &   0.031\\
 2.474 &  0.15 &  0.270 &   0.035\\
 2.624 &  0.15 &  0.327 &   0.040\\
 2.774 &  0.15 &  0.349 &   0.042\\
 2.924 &  0.15 &  0.378 &   0.046\\
 3.074 &  0.15 &  0.420 &   0.051\\
 3.224 &  0.15 &  0.388 &   0.054\\
 3.374 &  0.15 &  0.439 &   0.061\\
 3.524 &  0.15 &  0.569 &   0.068\\
 3.749 &  0.30 &  0.726 &   0.054\\
 3.974 &  0.15 &  0.807 &   0.084\\
 4.124 &  0.15 &  0.861 &   0.093\\
 4.274 &  0.15 &  0.967 &   0.095\\
 4.483 &  0.27 &  1.093 &   0.099\\
\noalign{\smallskip}\hline\noalign{\smallskip} 
\end{tabular}
\end{table}

\section{Evolution of the Ly$\alpha$ effective optical depth}\label{txt:tau_evol}

The \ion{H}{i} effective optical depth is related to the mean transmission as
$\tau_{\rm eff} \equiv -\ln\left <T\right> \equiv -\ln \left <
e^{-\tau_\ion{H}{i}}\right >$, where $\left < \: \right>$ indicates the
average over predefined redshift intervals. Note that the effective optical depth is defined as the
negative logarithm of the average pixel by pixel transmission, and not as the
average value of the pixel by pixel optical depth in a given redshift
interval. Its redshift evolution \citep{Zuo93b} is usually well represented by
a power law
\begin{equation}
\tau_{\rm{eff}}= \tau_0 (1+z)^{\gamma+1}.\label{eq:tau}
\end{equation}
We explored two methods of determining the evolution of $\tau_{\rm eff}$ with redshift.

In the first approach each Ly$\alpha$ forest spectrum was divided into two
redshift intervals, each covering the same redshift path length to sample the
 increasing opacity along single sight lines. The upper panel of
Fig.~\ref{fig:tauevol} shows the results obtained using all those pixels in
the Ly$\alpha$ forest range decontaminated from damped Ly$\alpha$ or Lyman
limit systems (DLAs and LLSs respectively) and the proximity effect region
($\Delta v<5000$km s$^{-1}$ were excluded).

The second approach consists of measuring the mean $\tau_{\rm eff}$ in the
forest of all quasars intersecting a particular redshift slice ($\Delta z =
0.15$). The results are shown in Table~\ref{tab:tau_comb} and in the lower panel of Fig.~\ref{fig:tauevol}, 
yielding best-fit values of $(\log\tau_0,\gamma) =(-2.21 \pm 0.09,2.04 \pm 0.17)$.

For both approaches the uncertainties in the effective optical depth,
$\sigma_{\tau_\mathrm{eff}}$, quantify the amount of variance between the
individual lines of sight for the considered redshift range. 

In Figure~\ref{fig:tauevol} we present the results from the two approaches in
comparison to the evolution of the Ly$\alpha$ optical depth as measured by
\citet{schaye03}, using a similar technique on 19 QSO spectra taken with UVES
and HIRES and the fit by \citet{kim07} combining 18 UVES spectra.  Our results
agree well with both \citet{schaye03} and \citet{kim07} fits. There is less
agreement at high redshift, probably due to the limited number of QSOs at
$z>3.6$, but our results are formally consistent within the errors.

Although both panels use the same data, these were combined in different ways
and thus return slightly different results. It is clear from the top panel
that the first method suffers from a lack of data points in the range
$3.6<z<4$, leading us to slightly underestimate the effective optical depth at
high redshift. The second approach significantly reduces the scatter caused by
cosmic variance as already pointed out by \citet{kim02}. It is also in good
agreement with literature estimates. Therefore we adopt the best-fit
parameters from the second method to describe the redshift dependence of the
normalised effective optical depth in our quasar spectra.

In the bottom panel of Fig.~\ref{fig:tauevol} we identify a marginally
significant departure of $\tau_{\rm eff}$ from the power law evolution in the
redshift range $3.2 \lesssim z \lesssim 3.4$. This feature, first discovered
by \citet{bernardi03} using SDSS quasar spectra (but see \citealt{mcdonald05}),
has been been interpreted to be a result of one or more of the following effects:
(i) a change in the IGM temperature; (ii) a change in the free electron number
density; or (iii) an enhancement in the UV background possibly connected to
 \ion{He}{ii} reionisation. Recently, \citet{giguere07} found the same dip
in $\tau_{\rm eff}(z)$ using a sample of ESI and HIRES spectra. In the light
of this debate and using a completely independent data sample, we confirm the
detection of this feature. We note, however, that its significance is only
2.6$\sigma$ (two pixels with each 1.8$\sigma$), so the detection can be
called no more than tentative.

\section{Methods of quantifying the proximity effect}\label{txt:method}

The direct signature of an enhanced ionisation field near bright quasars is a
reduction of the neutral hydrogen fraction on a physical scale of several
Mpc. Two principal techniques to detect it have been developed in the last
two decades, which we briefly review in the following subsections.

\subsection{Flux transmission statistics}\label{txt:flux_stat}

The method used in this work to reveal the proximity effect along single lines
of sight is based on the comparison between the observed effective optical
depth and the expected one in the Ly$\alpha$ forest
\citep{liske01}. Approaching the quasar systemic redshift, the photoionisation
rate of the source starts to dominate over the UVB. This effect leads to an
enhancement in H$^+$ relative to H$^0$, reducing the opacity of
absorbers close to the QSOs. This influence modifies the effective optical
depth, which becomes
\begin{equation}
\tau_\mathrm{eff}= \tau_0  (1+z)^{\gamma+1}(1+\omega)^{1-\beta}\label{eq:tauPE}
\end{equation}
\citep{liske01} where $\beta$ is the slope in the column density distribution
and $\omega$ is the ratio between the quasar and background photoionisation
rates.

Following the assumptions outlined in \citet{bajtlik88} to compute $\omega$, we
get
\begin{equation}
\omega(z)=\frac{f_{\nu}(\lambda_0(1+z) )}{4\pi J_{\nu_0}}\frac{1}{(1+z)}  
        \left(\frac{d_{L}(z_{\mathrm{q}},0)}{d_{L}(z_{\mathrm{q}},z)}\right)^{2}\label{eq:omega}
\end{equation}
with $z$ as the redshift along the LOS such that $z<z_{\mathrm{q}}$,
$d_{L}(z_{\mathrm{q}},0)$ the luminosity distance of the QSO to the observer,
and $d_{L}(z_{\mathrm{q}},z)$ as seen at any foreground redshift in the LOS.

Equation \ref{eq:omega} implies that the spectral shape of the QSO and the background are
approximately the same around 1~Ryd, which might not be exactely true. Assuming a quasar spectral 
shape $f_\nu\propto \nu^\alpha$, with $\alpha_\mathrm{q}\simeq -0.5$ \citep{berk01} and taking a background spectral index of $\alpha_\mathrm{b}\simeq -1.4$ \citep{agafonova05}, the resulting scale of $\omega$ is shifted by $\sim 0.1$ dex. However, there may be an indication of a break in the 
spectral slope of quasars around the Lyman limit \citep{telfer02}. Such a break would make the quasar spectrum more similar again to that of the background, reducing the effect on $\omega$.

We adopt the following notation: The ratio of the observed
optical depth to the one expected in the Ly$\alpha$ forest, or the
\emph{normalised effective optical depth} $\xi$, is given by
\begin{equation}
\xi=\frac{\tau_\mathrm{eff}}{\tau_0 (1+z)^{\gamma+1}}=(1+\omega)^{1-\beta}. \label{eq:xi}
\end{equation}
with the parameters $(\log\tau_0,\gamma)$ measured in
Sect.~\ref{txt:tau_evol}.

The proximity effect is apparent as a decrease in the normalised optical
depth below unity as $\omega \rightarrow \infty$, i.e. towards the quasar. We
make use of this technique, first to detect the proximity effect on our
combined sample of quasars (Sect.~\ref{txt:comb_PE}), then towards single
lines of sight (Sect.~\ref{txt:sglos}) and finally to assess the dependence of
the measured UV background on redshift and overdensities (Sect.~\ref{txt:uvb}).

\begin{figure}
\resizebox{\hsize}{!}{\includegraphics{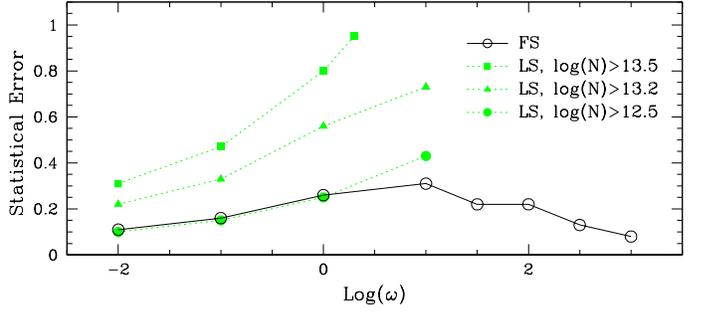}}
\caption{Statistical errors associated with individual lines of sight using the
line statistic (LS) and the flux statistic (FS) method. The different column
density thresholds represent the completeness limits from \citet{scott00} and
\citet{hu95} (solid squares and triangles, respectively). The filled circles
show the level of uncertainty if the line list were complete up to the minimum
column density at our resolution and S/N.}
\label{fig:ts_ls}
\end{figure}

\subsection{Line counting}\label{txt:line_stat}

Counting absorption lines stronger than a given threshold is historically the
first technique used to measure the deficit of absorption near high
redshift quasars. Typically represented by a Voigt profile, each absorption
line has to be fitted and associated with a set of parameters
$(N_{\ion{H}{i}},z,b)$ describing its column density, redshift, and Doppler
parameter, respectively. At medium resolution only the equivalent width can be
measured. Several authors \citep[e.g.,][]{giallongo96,scott00} used this
approach to detect the proximity effect and to measure the UV background on
combined samples of QSOs with the formalism introduced by \citet{bajtlik88}.
It is clear that at low spectral resolution, line counting is vastly inferior
to the flux transmisson statistic because of line blending.  On the other
hand, at the high resolution of UVES and similar instruments, line counting
could still be a reasonable approach. We therefore conducted a performance
comparison of the two techniques, following the prescription in
Sect.~\ref{txt:sims}. We simulated the Ly$\alpha$ forest of a quasar at
redshift $z=3$, obtaining not only the spectrum but also keeping the
\emph{complete} line list. We repeated this step to create 500 independent
simulated spectra.

The principal limitation of the absorption line fitting approach is line
blending, leading to incompleteness of weak lines at low column
densities. Even at high resolution where lines with $\log N_{\ion{H}{i}}
\simeq 12.5$ are detectable, the completeness is $<25\%$ for column densities
$\log N_{\ion{H}{i}} < 13.2$ \citep{hu95}. Another problem in fitting
absorption lines is that there is no unique solution, and much
is left to the experience and ability of the scientist/software code in
finding a \emph{good} fit. For our simulation study we employed the code
AUTOVP\footnote{Developed by R. Dav\'{e}: http://ursa.as.arizona.edu/$\sim$rad} to
fully automatically fit an array of absorption lines to each spectrum.
The proximity effect signature is then a decrease in the number density 
of lines $N$ relative to that expected for the redshift. Thus we
can define an alternative quantity 
\begin{equation}
\xi'=\frac{N}{N_0} = (1+\omega)^{1-\beta}, \label{eq:xi2}
\end{equation}
which again is expected to decrease from unity to zero as $\omega$ increases.
We measured $\xi$ (using the flux transmission statistic) and $\xi'$ (from the
automated line counts, adopting different column density thresholds) for all
500 simulated spectra, and computed mean and dispersion values for several
bins of $\log\omega$.

Figure~\ref{fig:ts_ls} presents the results of this study, showing the
standard deviation of $\xi$ or $\xi'$, i.e.\ the statistical error of the
measurement for a single line of sight, as a function of $\omega$ (approaching
the quasar). For a realistic completeness limit in column density
$\log N_\ion{H}{i}>13$, the flux transmission approach is always more sensitive
than line counting. In the unrealistic ideal case where we used the true input
line lists, the two methods are equivalent at low $\omega$, but even then the
line counting approach fails in probing $\xi'$ very close to the emission
redshift of the quasar. This can easily be understood as the quasar radiation
reduces the column densities by a factor $1+\omega$; thus, close to the quasar,
where $\omega \ga 100$, original column densities of $\log N_{\ion{H}{i}} \ga
15$ are needed to produce a detectable absorption line. Such absorbers are
very rare, which leads to increased uncertainties.

We conclude that line counting is less sensitive to the expected proximity
effect signature than the flux transmission method, even at high spectral
resolution, and all our results will be based exclusively on the flux
transmission technique.

\begin{figure}
\resizebox{\hsize}{!}{\includegraphics{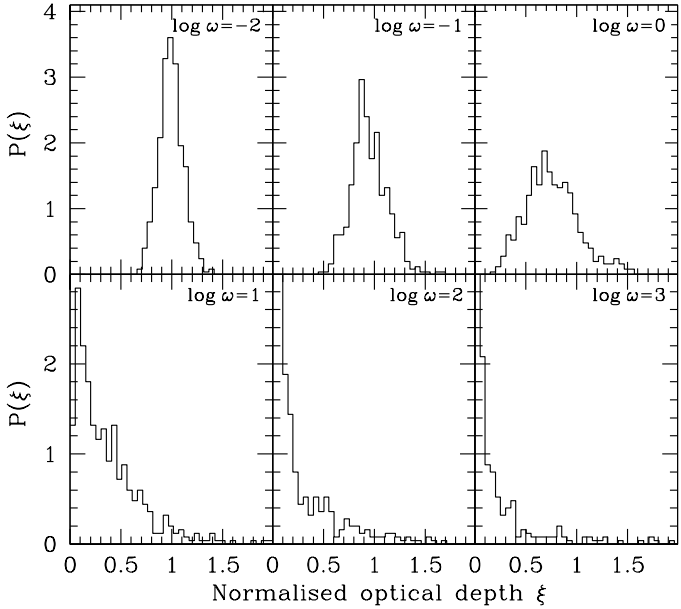}}
\caption{Distribution of the normalised optical depth $\xi$ under the
influence of increasing $\omega$ values. The histograms represent the $\xi$
distribution of 500 simulated spectra with $z_\mathrm{QSO} = 3$.}
\label{fig:pdf_xi}
\end{figure}

\subsection{Poissonian variance between lines of sight}\label{txt:errors}

At the spectral resolution in our sample, the major source of uncertainty is
due to the variation in the number of absorption features along individual
lines of sight. 
The probability of finding or not finding an absorption line at a given redshift 
is to first  approximation a Poissonian process which 
can be easily modelled by our Monte Carlo simulations. While a full accounting
of the expected effects of cosmic variance due to the large-scale structure
of the IGM would require cosmological 3D simulations, it is actually useful
to start with the Poissonian approximation, as it allows us to appreciate to what
extent the observed level of variance is consistent with pure Poisson noise.

Equation \ref{eq:omega} shows that $\omega$ depends on the redshift and
luminosity of the quasar, implying that each QSO will have its own particular
$\log \omega$ scale. Consequently, in a fixed $\log \omega$ interval the
influence of Poisson noise will vary from line of sight to line of sight,
since the corresponding redshift range in the Ly$\alpha$ forest is determined
by the quasar emission redshift and its Lyman limit luminosity. Therefore, the
distribution of the normalised optical depth $\xi$ as a function of $\log \omega
$ will depend on (i) Poisson noise, (ii) quasar redshift, and (iii) quasar
luminosity.

We studied how the distribution of $\xi$ is affected by different UV radiation
fields (i.e. in different $\log \omega $ bins) at a given redshift. We
computed the standard deviation $\sigma$ given by
\begin{equation}
\sigma^2=\int (\xi-\mu)^2\ P(\xi)\, \mathrm{d}\xi
\end{equation}
where $P(\xi)$ describes the probability density function of $\xi$ estimated
using our mock catalogue of sight lines as displayed in Fig.~\ref{fig:pdf_xi},
and $\mu$ is its mean value. For illustration purposes, we assumed a quasar at
redshift $z=3$ and luminosity $\log L_{\nu_0}=31.5$ with a binning of $\Delta
\log \omega =1$. Each panel displays how Poisson noise affects a different
portion of the Ly$\alpha$ forest according to the quasar properties, i.e.\ for
different values of $\omega$.  The final error bars in the proximity effect
analysis are a composition of the following three effects, ordered by
descending relative importance: (i) the standard deviation computed above,
(ii) the continuum uncertainties (Sect.~\ref{txt:cont}) and (iii) the
intrinsic noise present in the spectrum.  The Poisson noise in the placement
of lines dominates the total error. Continuum uncertainties and the intrinsic
noise are almost negligible.

\begin{figure}
\resizebox{\hsize}{!}{\includegraphics*{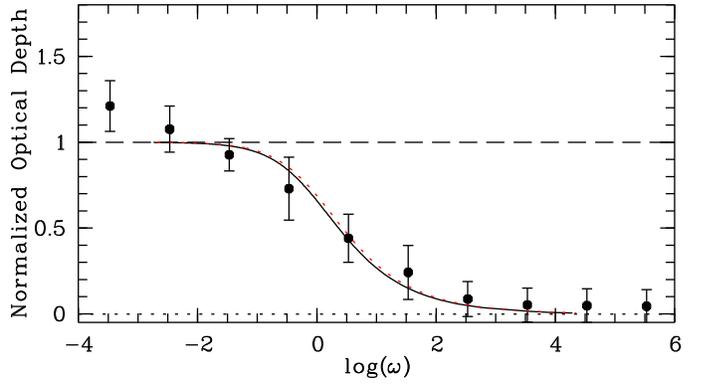}}
\caption{Normalised effective optical depth versus $\omega$ for the combined
sample of 40 quasars, binned in steps of $\Delta\log\omega = 1$. The curved
lines are the best fit of the simple photoionisation model to the data,
corresponding to a UV background of $\log(J_{\nu_0})=-21.10$ (solid line) in
units of erg~cm$^{-2}$~s$^{-1}$~Hz$^{-1}$~sr$^{-1}$, while the dotted curve
denotes to the fiducial UVB ($J^\star$) for the computation of $\omega$. The
horizontal dashed line illustrates the case of no proximity effect.}
\label{fig:PE_combined}
\end{figure}

\begin{figure*}
\resizebox{\hsize}{!}{\includegraphics{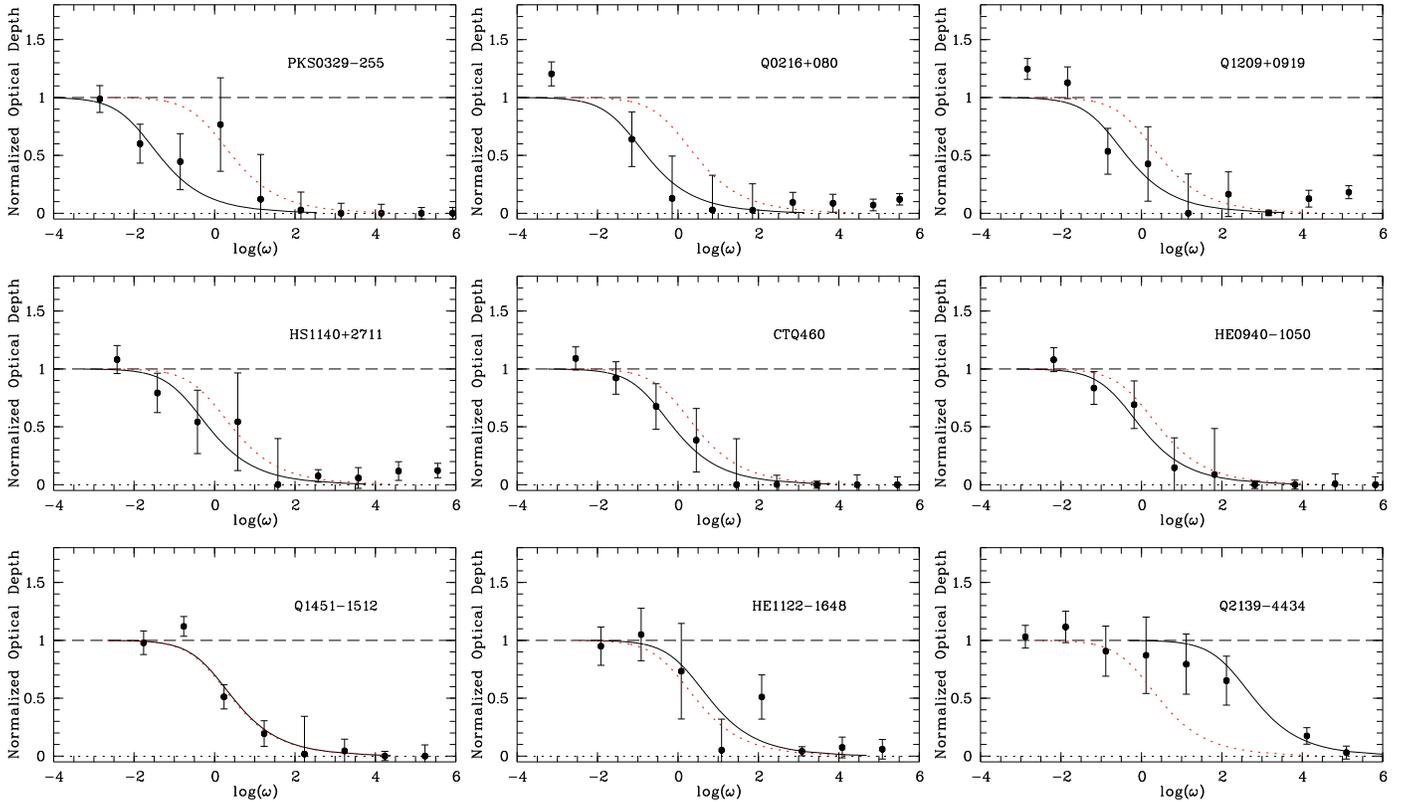}}
\caption{The proximity effect signatures in individual lines of sight, for a
subsample of 9 QSOs.  Each panel shows the normalised effective optical depth
$\xi$ versus $\omega$, binned in steps of $\Delta\log\omega = 1$, with the
best-fit model of the combined analysis superimposed as the dotted red line
(see Sect.~\ref{txt:comb_PE}). The solid lines delineate the best fit to each
individual QSO as described in the text. This subsample was chosen for
presentation purposes to show the variable strength of the proximity effect,
going from \emph{strong} (top-left panel) to \emph{weak} (bottom-right
panel).}
\label{fig:PE_prof_sample}
\end{figure*}

\section{The proximity effect in the combined sample of quasars}\label{txt:comb_PE}

We first present briefly the evidence of the proximity effect in the traditional way,
as a signal in the combined full sample of QSOs. The result is displayed in 
Fig.~\ref{fig:PE_combined}. As expected, the normalised effective optical
depth $\xi$ starts at unity for $\omega\ll 1$ and then goes down to zero for 
$\omega\gg 1$. In the definition of $\omega$ (Eq.~\ref{eq:omega}) 
we had to assume a fiducial value of the UVB intensity, $J^\star = 10^{-21}\,
\mathrm{erg}\,\mathrm{cm}^{-2}\,\mathrm{s}^{-1}\,\mathrm{Hz}^{-1}\,\mathrm{sr}^{-1}$.
This uniquely converts the redshift scale into an $\omega$ scale. We defined a regularly spaced
grid along the $\log \omega$ axis and in each bin we determined the average transmission and effective optical depth values, and using Eq.~\ref{eq:xi} the corresponding values of $\xi$.
We then modelled the decrease in $\xi$ with $\log(\omega)$ according to the formula
\begin{equation}
F(\omega)=\left(1+\frac{\omega}{a}\right)^{1-\beta}\label{eq:fit}
\end{equation}
where $a$ is the single free parameter which expresses the \emph{observed}
turnover of $\xi$ and thereby provides the best-fit value of $J_{\nu_0}$, since
obviously $J_{\nu_0} = a\times J^\star$. The slope of the column density 
distribution was fixed to $\beta = 1.5$ (see Sect.~\ref{txt:sims} for details).
 
For the median redshift of our sample of $z = 2.73$ we thus obtain $J_{\nu_0}
= (7.9 \pm 3.1) \times
10^{-22}\,\mathrm{erg}\,\mathrm{cm}^{-2}\,\mathrm{s}^{-1}\,\mathrm{Hz}^{-1}\,\mathrm{sr}^{-1}$,
or in logarithmic units $\log J_{\nu_0} = -21.10^{+0.14}_{-0.22}$.  This
number is very close to the value of $\log J_{\nu_0} = -21.03$ reported by us
in Paper~I and is also consistent with several other literature estimates. A
detailed comparison with the literature is hampered by the fact that the vast majority of
previous UVB measurements were derived adopting the almost obsolete
Einstein-de~Sitter model. Typically, values for $J_{\nu_0}$ are then about a factor
of 1.4 (0.15~dex) higher than in a $\Lambda-$Universe.  \citet{cooke97}
measured $\log J_{\nu_0} = -21.00_{-0.15}^{+0.18}$ on 11 high resolution quasar spectra
using line statistics.  \citet{scott00} obtained $\log J_{\nu_0} =
-21.15_{-0.43}^{+0.17}$, applying the same method on more than a hundred spectra
at $\sim 1$~\AA\ resolution. \citet{liske01}
used the flux statistic on 10 QSO spectra with $\sim 2$~\AA\ resolution and a
S/N of $\sim 40$, obtaining $\log J_{\nu_0} = -21.45_{-0.20}^{+0.30}$.

While the formal errors of this measurement are small, there are reasons to
believe that the proximity effect exploited this way delivers systematically
too high values of $J_{\nu_0}$. Overdense environments around quasars have 
been suggested as the prime reason; we shall demonstrate in the next section
that in fact the averaging process inherent in the sample combination is
responsible for a major bias in the UVB determination.

\begin{figure}
\resizebox{\hsize}{!}{\includegraphics{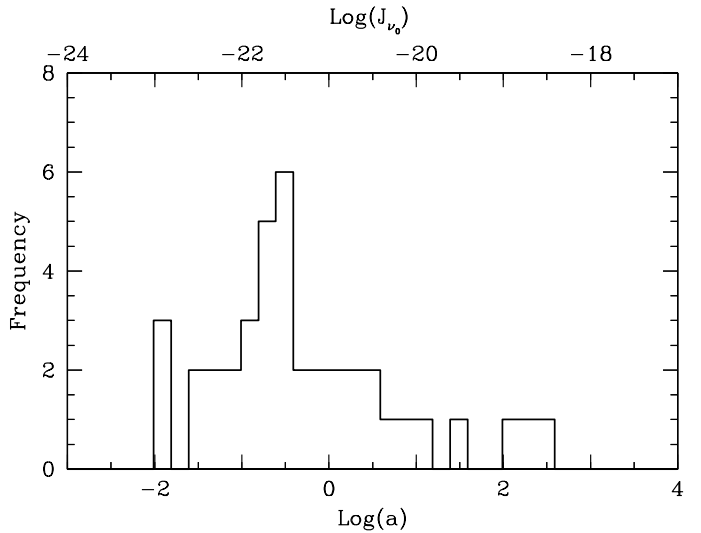}}
\caption{Observed distribution of the proximity effect strength (given by
$\log a$) for our sample of objects. The distribution is characterised by a
prominent peak and a skewed profile which extends over several orders of
magnitude in strength.}
\label{fig:log_a_data1}
\end{figure}

\begin{table}
\caption{The proximity effect strength $\log a$ along single lines of sight.}
\label{tab:log_a}
\centering
\begin{tabular}{lccc}
\hline\hline\noalign{\smallskip}
QSO & $z_{\rm{em}}$ & $\log L_{\nu_0}^\dagger$ & $\log a $\\
\noalign{\smallskip}\hline\noalign{\smallskip}
Q~1232$+$0815   & 2.575  & 30.91 & $-1.89 \pm 0.37$ \\ 
Q~0329$-$385    & 2.438  & 31.48 & $-1.86 \pm 0.29$ \\ 
PKS~0329$-$255  & 2.706  & 31.58 & $-1.86 \pm 0.25$ \\ 
HE~0001$-$2340  & 2.278  & 31.68 & $-1.46 \pm 0.36$ \\ 
HE~1347$-$2457  & 2.615  & 32.06 & $-1.43 \pm 0.25$ \\ 
Q~0216$+$080    & 2.996  & 31.20 & $-1.26 \pm 0.36$ \\ 
HE~2217$-$2818  & 2.414  & 32.02 & $-1.21 \pm 0.31$ \\ 
Q~0109$-$3518   & 2.406  & 31.72 & $-1.10 \pm 0.38$ \\ 
Q~1249$-$0159   & 3.668  & 31.27 & $-1.08 \pm 0.22$ \\ 
PKS~2126$-$158  & 3.285  & 31.92 & $-0.99 \pm 0.22$ \\ 
Q~1209$+$0919   & 3.291  & 31.42 & $-0.85 \pm 0.29$ \\ 
Q~0913$+$0715   & 2.788  & 31.48 & $-0.84 \pm 0.34$ \\ 
Q~0002$-$422    & 2.769  & 31.76 & $-0.78 \pm 0.32$ \\ 
PKS~1448$-$232  & 2.222  & 31.50 & $-0.75 \pm 0.47$ \\ 
Q~0055$-$2659   & 3.665  & 31.81 & $-0.67 \pm 0.20$ \\ 
HS~1140$+$2711  & 2.628  & 31.94 & $-0.63 \pm 0.35$ \\ 
HE~2243$-$6031  & 3.012  & 32.17 & $-0.61 \pm 0.24$ \\ 
CTQ~0460        & 3.141  & 31.76 & $-0.61 \pm 0.27$ \\ 
HE~0940$-$1050  & 3.089  & 32.13 & $-0.49 \pm 0.26$ \\ 
Q~1621$-$0042   & 3.709  & 31.91 & $-0.49 \pm 0.20$ \\ 
Q~1151$+$0651   & 2.758  & 31.21 & $-0.45 \pm 0.46$ \\ 
PKS~2000$-$330  & 3.786  & 31.69 & $-0.45 \pm 0.23$ \\ 
Q~0122$-$380    & 2.192  & 31.51 & $-0.42 \pm 0.44$ \\ 
Q~1409$+$095    & 2.843  & 31.06 & $-0.33 \pm 0.54$ \\ 
Q~0347$-$3819   & 3.229  & 31.66 & $-0.23 \pm 0.30$ \\ 
Q~0420$-$388    & 3.120  & 31.93 & $-0.18 \pm 0.32$ \\ 
Q~1451$-$1512   & 4.766  & 32.12 & $-0.01 \pm 0.14$ \\ 
Q~1317$-$0507   & 3.719  & 31.35 & $+0.13 \pm 0.31$ \\
Q~0453$-$423    & 2.664  & 31.41 & $+0.18 \pm 0.45$ \\ 
HE~1122$-$1648  & 2.407  & 32.05 & $+0.30 \pm 0.33$ \\ 
HE~1158$-$1843  & 2.459  & 31.56 & $+0.35 \pm 0.38$ \\ 
BR~1202$-$0725  & 4.697  & 32.24 & $+0.51 \pm 0.14$ \\
Q~1223$+$1753   & 2.955  & 31.16 & $+0.58 \pm 0.36$ \\ 
CTQ~0247        & 3.025  & 31.66 & $+0.76 \pm 0.25$ \\ 
HE~2347$-$4342  & 2.886  & 32.05 & $+0.87 \pm 0.17$ \\ 
Q~2206$-$1958   & 2.567  & 31.57 & $+1.15 \pm 0.21$ \\ 
HE~1341$-$1020  & 2.137  & 31.17 & $+1.51 \pm 0.14$ \\ 
PKS~0237$-$233   & 2.224  & 31.93 & $+2.07 \pm 0.12$ \\ 
Q~2139$-$4434   & 3.207  & 31.39 & $+2.32 \pm 0.26$ \\ 
HE~0151$-$4326  & 2.787  & 31.92 & $+2.57 \pm 0.15$ \\ 
\noalign{\smallskip}\hline\noalign{\smallskip}
\end{tabular}                              
\begin{list}{}{}
\item[$\dagger$:] Lyman limit luminosities and uncertainties 
$\left( \sigma_{\log L_{\nu_0}}\simeq^{+0.05}_{-0.1} \right)$ in units of 
$\mathrm{erg}\,\mathrm{s}^{-1}\,\mathrm{Hz}^{-1}$
\end{list}
\end{table}

\section{The proximity effect in individual lines of sight}\label{txt:sglos}

\begin{figure*}
\includegraphics[width=\textwidth]{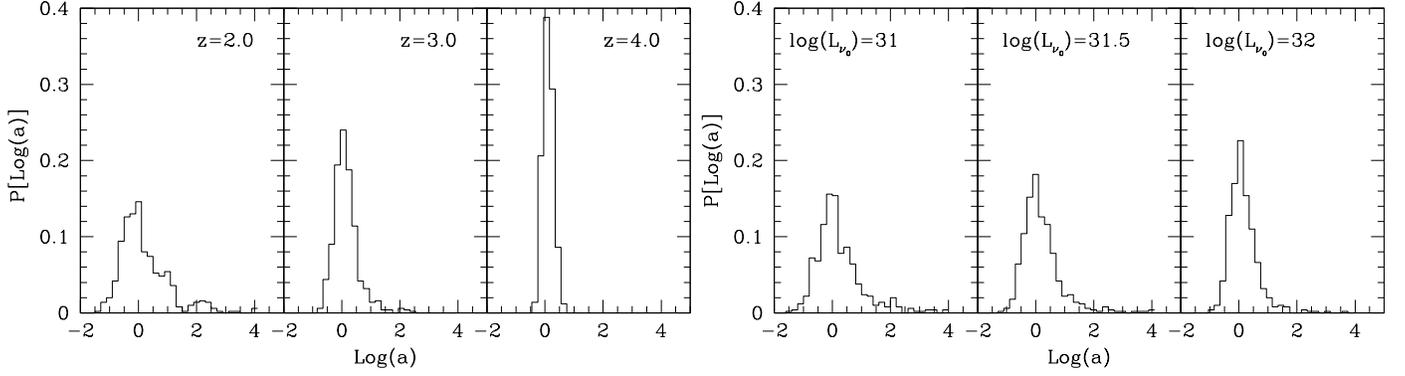}
\caption{Proximity effect strength distributions in our synthetic spectra for
different quasar redshifts (left-hand panel) and luminosities (right-hand
panel). The distributions are computed from 500 artificial sight lines
generated by our Monte Carlo code as described in Sect.~\ref{txt:sims} and are
affected only by Poissonian shot noise.}
\label{fig:log_a_distMC}
\end{figure*}

\subsection{The distribution of the proximity effect strengths}\label{txt:pepd}

For our investigation of the proximity effect towards single lines of sight,
the approach was essentially identical to that of the combined analysis: We
computed the normalised effective optical depths for individual sight lines
within a given range $\Delta\log\omega$ and checked whether $\xi$
systematically decreases for high values of $\omega$. For presentation
purposes, the results of a subsample of 9 objects are displayed in
Fig.~\ref{fig:PE_prof_sample}, one panel per quasar (see
Appendix~\ref{app:PElos} for the complete sample).  The error bars are
dominated by Poissonian shot noise, estimated from the simulations as
described in Sect.~\ref{txt:errors}. In each panel, $\xi(\omega)$ is shown for
the fiducial reference value $J_{\nu_0} = J^\star $.  Figures~\ref{fig:PE_prof_sample} and
\ref{fig:PE_prof_sample1} illustrate that in all cases, $\xi$ decreases
substantially as $\omega$ increases. Thus we conclude that the proximity
effect is detected in all of our quasar spectra. This is in excellent agreement
with our finding in Paper~I, where in only 1 out of 17 low-resolution quasar spectra 
we failed to detect the proximity effect  (and for that object,
HE~2347$-$4342, we now detect a weak effect in the high-resolution UVES spectrum;
see Appendix~\ref{app:PElos}).

We then applied the above fitting procedure (Eq.\ref{eq:fit}) to each spectrum
separately. The results are shown as the solid curves in
Fig.~\ref{fig:PE_prof_sample}, while Table~\ref{tab:log_a} summarises the fit
results.  The value of the fitting parameter $a$, which describes the
horizontal offset of the solid curve relative to the dotted one, will be
regarded in the following as a measure of the strength of the proximity effect
signal. It shows significant scatter between the different
quasars. Note that the zeropoint of $\log a$ is arbitrary and depends only on
our assumed fiducial UVB value $J^\star$. 
Figure~\ref{fig:log_a_data1} presents the measured proximity effect
strength distribution (PESD), i.e.\ the distribution of $\log a$ for our sample.

Three major characteristics emerge: \\
(i) the distribution covers almost five
orders of magnitude in $a$, ranging from a very weak proximity effect ($a\gg
1$) to a very strong one ($a\ll 1$), \\
(ii) it shows a well-defined peak, and \\
(iii) it is significantly skewed. \\
In the following we will investigate the effects behind these properties.

\subsection{The effects of variance between sight lines}\label{txt:ll_z_dep}

The most obvious contributor to a spread in proximity effect strengths is the
variance in the absorber distribution between different lines of sight. To
investigate this effect we employed our Monte Carlo simulated spectra, for 
which by design this variance can be accurately described as a pure Poissonian
process. 

The proximity effect was introduced in the synthetic spectra as a reduction in
the optical depth by a factor $1+\omega$, with $\omega$ given by
Eq.~\ref{eq:omega}. We first fixed the
luminosity of the quasar to a value of $\log(L_{\nu_0}/\mathrm{erg\ s^{-1}\
Hz^{-1}})=31.5$ and considered three different redshifts: $z$ = 2.0, 3.0, and
4.0, in order to see how the distribution of $a$ changes with $z$. Then we took a
constant redshift of $z=3.0$ and considered three quasar luminosities:
$\log(L_{\nu_0}/\mathrm{erg\ s^{-1}\ Hz^{-1}})$ = 31, 31.5, and 32. The UV
background was fixed to $J_{\nu_0}=J^\star$ in both cases. Our results are presented
in Fig.~\ref{fig:log_a_distMC}; recall that high $a$ values imply a weak proximity
effect.

Two main characteristics become apparent in Fig.~\ref{fig:log_a_distMC}: The
distributions are significantly skewed, and the skewness decreases towards
higher redshifts and higher luminosities. This asymmetry is a direct result 
of the non-linear dependence of $\omega$ on redshift as expressed in eq.~\ref{eq:omega}. 
Towards the quasar redshift equal $\Delta \log \omega$ bins correspond to progressively
smaller $\Delta z$ intervals. Very close to the QSO, the (Poissonian) distribution of 
the number of absorbers per  $\Delta \log \omega$ bin deviates significantly from a Gaussian. 
Similarly the distribution of effective optical depth becomes skewed at 
$z\rightarrow z_\mathrm{em}$, resulting in  high $\tau_\mathrm{eff}$ values being more 
likely than expected for a normal distribution. The asymmetry of the PESD directly reflects 
this statistical effect.

  The dependence with redshift can be explained by the fact that on
the one hand, Poisson noise has a weaker impact at high redshifts and on the other
hand, an increasing luminosity is capable of a substantial reduction in the
optical depth further away from the quasar emission, overionising even strong
systems in the quasar vicinity. 

Because of the asymmetric nature of the PESD at any given redshift and
luminosity, one tends to overestimate the best fit values of Eq.~\ref{eq:fit}
when combining multiple sight lines together in an averaging process. By
comparing how the distribution widths evolve with redshift and
luminosity, we conclude that luminosity has a weaker effect than redshift in
changing the shape of the PESD. 

A remarkable feature of Fig.~\ref{fig:log_a_distMC} is the fact that the
mode of the distribution always stays at the input value corresponding
to $\log a = 0$. We exploit this property in the next section.

\begin{figure}
\resizebox{\hsize}{!}{\includegraphics{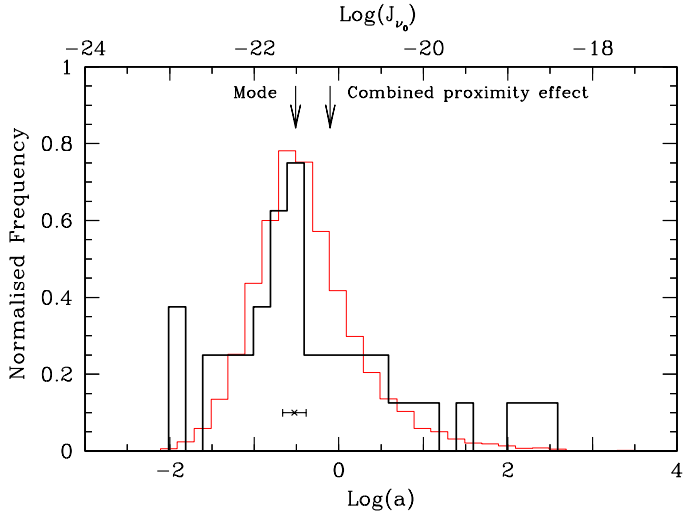}}
\caption{Comparison between observed (thick) and simulated (thin) distribution of the proximity effect 
strength (given by $\log(a)$) for our sample of quasars. We mark the position of the mode of the observed 
distribution and the relative error associated to it. The error bars are computed from the dispersion of 
the mode in the 500 realisations of the simulated sample. We also display the value of the UVB computed in 
the combined analysis of the proximity effect.}
\label{fig:log_a_data}
\end{figure}

\subsection{An unbiased measurement of the UV background}\label{txt:unbiasuvb}

Unfortunately, we do not have 500 observed quasar spectra at one particular
redshift or luminosity, but rather a quasar sample distributed in redshift and
luminosity. We created another set of simulated spectra to predict the `PESD as observed',
assuming that only Poissonian variance contributes to the spread of $\log a$,
in the following way: We constructed 40 simulated QSO spectra with the condition
that their redshifts and luminosities were exactly the same
as the observed ones, and we constructed the PESD for this `sample'. We
repeated the process 500 times and averaged the PESD. The result is shown
in Fig.~\ref{fig:log_a_data} as a thin histogram, superimposed on the observed
data of our quasar sample. 

The predicted distribution of the proximity effect strength parameter $\log a$ 
looks amazingly similar to the observed one. In particular, the degree of
asymmetry is very nearly the same. The observed PESD seems to show a slightly
narrower core, and it certainly has some outliers on both sides, at 
$\log a \simeq -2$ and $\log a > 2$. We interpret this excellent
agreement as an indication that the statistical fluctuations of the
distribution of absorption features between individual sight lines is 
responsible for most of the spread and in fact for most of the asymmetry 
in the PESD.

What does this imply for the UV background? We again find that the modal value
of the `as observed' distribution recovers the input UVB intensity, whereas
the combined analysis heavily overestimates it. From the 500 realisations we
derive that the mode of the PESD of 40 quasars is consistent with the value of
input model to within $\pm 0.15$~dex. Similarly we can quantify the bias
introduced by the combined analysis. For this purpose, we computed the
best-fit $\log a$ as in Sect.~\ref{txt:comb_PE} and averaged it over the 500
realisations. Typically, the averaged value of $J_{\nu_0}$ is overestimated by
$\Delta \log a = 0.3$ dex, or a factor of 2.

\begin{figure}
\resizebox{\hsize}{!}{\includegraphics{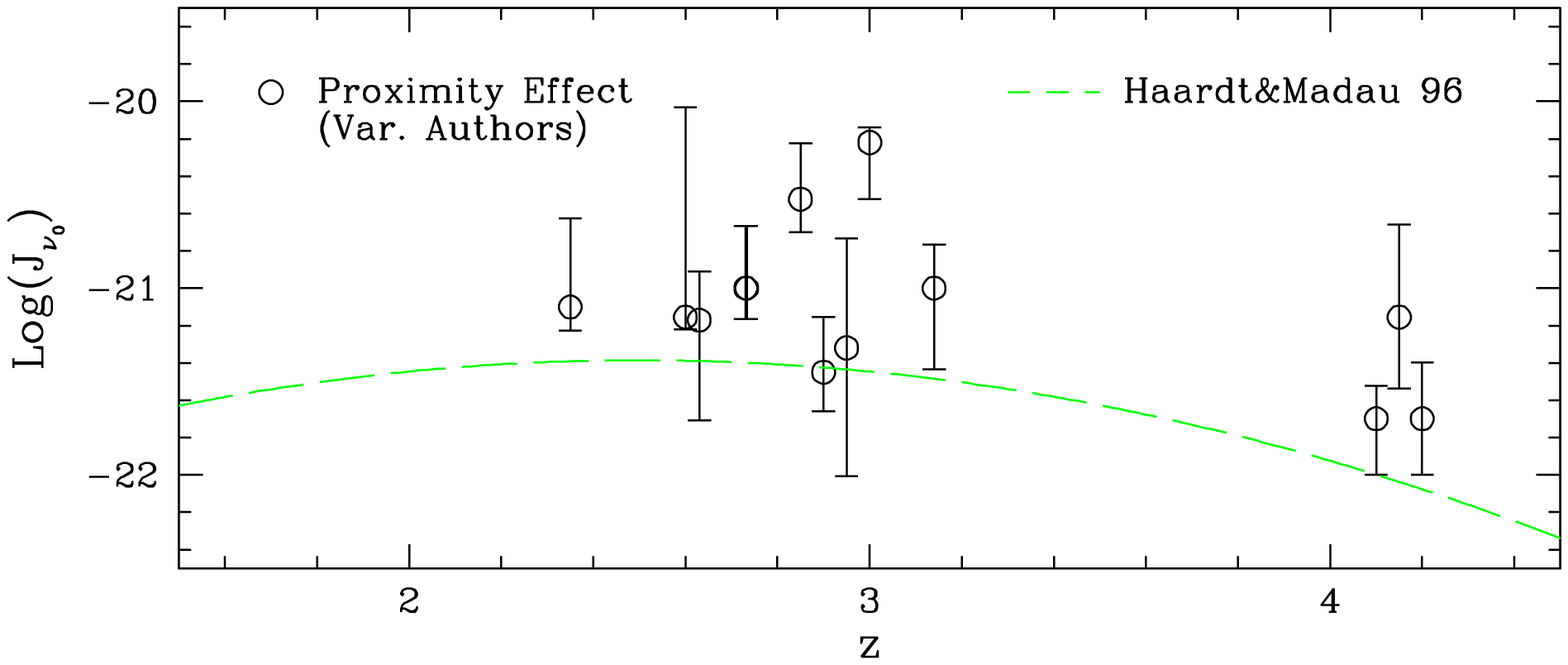}}
\caption{Redshift evolution of the UVB modelled by \citet{haardt96} in
comparison to previous measurements of the UVB via the proximity effect
\citep{bajtlik88,lu91,giallongo93,williger94,bechtold94,fernandez-soto95,srianand96,
giallongo96,lu96,savaglio97,cooke97,scott00,liske01},
all for an Einstein-de~Sitter universe. }
\label{fig:uvb_evollit}
\end{figure}

\begin{figure*}
\sidecaption
\includegraphics[width=12cm]{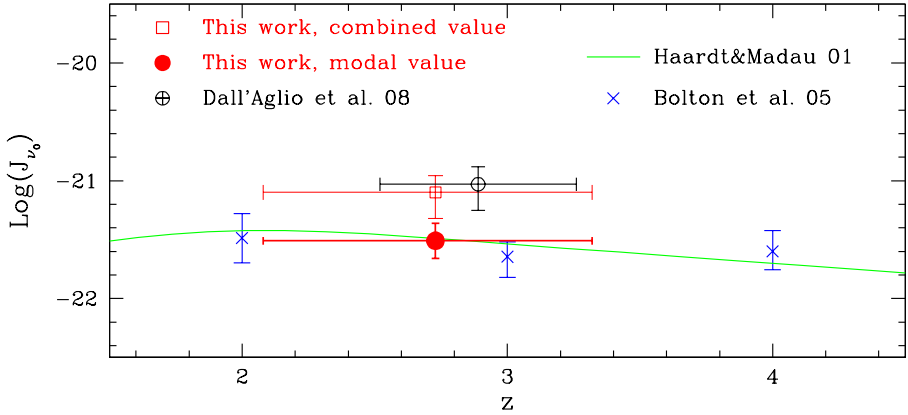}
\caption{Measured intensity of the UVB for a $\Lambda$ universe. Our best
  estimate is the modal value of $J_{\nu_0}$ (filled red circle); the 
  biased `combined value' (open square) as well as the similarly obtained
  value from Paper~I (open circle) are shown for comparison. Also shown 
  are the most recent evolving UVB model by \citet{haardt01} (continuous
  green line) and the determination by \citet{bolton05}. The horizontal 
  bars indicate the lower and upper quartile of the quasar redshifts.}
\label{fig:uvb_evolmode}
\end{figure*}

From Fig.~\ref{fig:log_a_data} we estimated the position of the mode of the
observed PESD to be located at $\log a = -0.51$ and converted it into a
measurement of the UVB, yielding a value of $\log J_{\nu_0} = -21.51 \pm 0.15$.  The uncertainty
was derived from the dispersion of the mode in the simulated PESDs. This value
of the UVB is 0.4~dex lower than the best-fit of the combined proximity effect
analysis given in Sect.~\ref{txt:comb_PE}.

For completeness we also considered what happens when one computes 
the average of the best-fit $\log a$ values for individual quasars (as listed
in Table~\ref{tab:log_a}), instead of fitting the average $\xi$-$\omega$ profile.
This results in a UVB value of $\log J_{\nu_0} = -21.24\pm0.17$, lower than
the outcome of the combined proximity effect analysis as well, but still much
higher than the modal value (and also systematically offset in the simulations).

We compiled several literature results in measuring the UVB by the proximity
effect together with our new our results in Figs.~\ref{fig:uvb_evollit} and 
\ref{fig:uvb_evolmode}. As said above, the cosmological model employed does
have an effect on the UVB result. Figure~\ref{fig:uvb_evollit} collects 
literature measurements that were performed for an Einstein-de~Sitter model. 
Most of them agree poorly with the UVB model by \citet{haardt96} (which 
was computed for the same cosmological model), in the sense that the
measurements are systematically higher.

Figure~\ref{fig:uvb_evolmode} puts our new determination of $J_{\nu_0}$ into
context. It is remarkable that our modal value is almost exactly on the 
predicted curve based on the newer UVB model by \citet{haardt01}.
Their computation was based on assuming a combined contribution of QSOs and 
star-forming galaxies to the UV background, now performed for a $\Lambda$ universe.
It is also fully consistent with the determination of the
UVB photoionisation rate by \citet{bolton05} obtained by matching the observed Ly 
forest transmission to cosmological simulations. It is equally evident
that our `combined value' is too high, as is our similarly obtained best-fit
value in Paper~I.

We summarise this section by stating that the distribution of proximity
effect strengths as determined in individual quasar spectra is largely
consistent with a purely Poissonian random process. We thus understand
why the PESD is asymmetric, and why previous measurements based on averaging
sample properties were bound to overestimate the UVB background. So far
we have identified the modal value of the PESD as a largely unbiased estimator
of the UVB. In the next section we will improve on this estimator by 
proposing a hybrid approach that also takes physical overdensities into
account.

\section{The role of overdensities around quasars}\label{txt:data_overd}

\subsection{Quantifying overdensities}\label{txt:xi_def}

We have demonstrated above that much of the disagreement between proximity
effect and other measurements of the UVB is due to the asymmetric shape of the
PESD, even for the idealised case of purely Poissonian variance between sight
lines. However, this does not imply that other effects can be neglected
altogether. The most suspicious additional bias in this context is the degree
of typical H$^0$ overdensity within a few Mpc around quasars. Obviously,
if a quasar environment is overdense and thus more likely to show high column
density absorption than the general Lyman forest, the proximity effect will
appear to be weakened \citep{loeb95}. 

This issue has been recently addressed by \citet{rollinde05} and
\citet{guimares07} using both observations and numerical simulations.  Their
basic approach consisted of a comparison of the cumulative optical depth
probability distributions at difference distances to their quasars.  Close to
the quasars, that distribution deviates from the one for the general Lyman
forest. Assuming that the UVB photoionisation rate is well constrained from
the cosmic mean H$^0$ opacity and accounting for the expected proximity effect
signature based on the known quasar luminosities, they concluded from the
discrepancy between the measured and predicted distributions that quasars
reside in physical overdensities of a factor of a few (averaged over $\sim
2.5h^{-1}$ proper Mpc). The amount of inferred overdensity depends,
as one would expect, on the adopted value of the general UVB photoionisation rate.

Here we approach the same problem from a different angle. We also assume that
the mean UVB photoionisation rate, or the mean intensity $J_{\nu_0}$, can be
measured. We adopt the modal value of the PESD estimated in
Sect.~\ref{txt:unbiasuvb}, which we demonstrated above to be in excellent
agreement with other methods of estimating the UVB. We then again consider each
individual quasar in our sample in turn, aiming at measuring the
\emph{distribution} of (over)density properties. For the present paper we are
not interested in estimating physical overdensities, but only in identifying the
quasars with strongly overdense environments. We therefore adopted a pragmatic
definition of `density' that simply involves the absorber optical depth within
a certain filter length as a proxy.  We computed this `overdensity degree'
$\Xi$ in the following way:

\begin{figure}
\resizebox{\hsize}{!}{\includegraphics{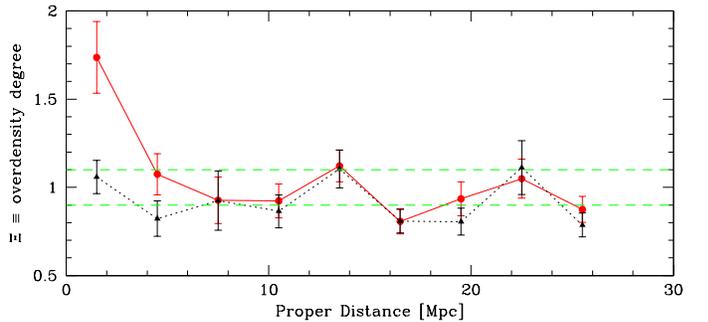}}
\caption{Average overdensity degree $\Xi$ as function of proper distance for the
complete sample of quasars (solid line) and in a subsample with $\Xi$
consistent with no sign of overdensities within 3~Mpc from the quasar (dotted
line). The two horizontal lines mark the $\pm 1\sigma$ scatter expected from
our simulation in the case of no overdense environment.}
\label{fig:xifluctuation}
\end{figure}

\begin{enumerate}
\item We reconstructed the optical depth profile $\tau^*(z)$, \emph{removing
the radiative influence of the quasar} by multiplying the optical depth
$\tau(z)$ in the Ly$\alpha$ forest by $1+\omega(z)$. For the UVB intensity we
used the modal value of the PESD, $\log J_{\nu_0} = -21.51$. Despite the high
S/N of our spectra, this procedure required particular attention for
transmission points falling below zero and above unity due to noise and
continuum uncertainties. The negative points belong to saturated lines which
will become even more saturated after the inverse correction for the proximity
effect; such points were left uncorrected. Pixels with transmission values
$T>1$ were also disregarded in the correction in order to prevent numerical
problems or an artificially increased S/N level. Both cases apply only to  
a small number of pixels, and their in- or exclusion does not affect the
results.

\begin{figure*}
\sidecaption
\includegraphics[width=12cm]{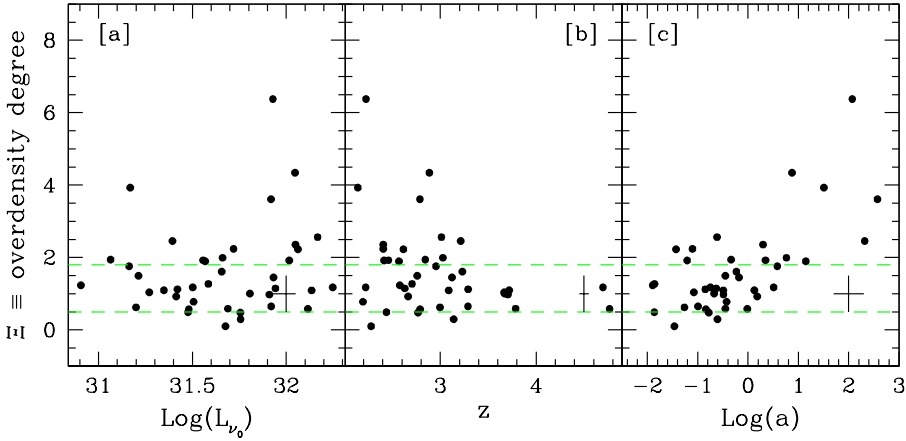}
\caption{Dependence of the overdensity degree on quasar luminosity (panel
[a]), redshift (panel [b]), and on the proximity effect strength (panel
[c]). The uncertainties relative to each measurements are represented by the
solid cross in the respective panel. The solid line in the right-hand panels
shows the least-square linear fit to the data. In each panel the dashed
horizontal lines give an estimate of the amount of Poissonian shot noise in
the simulations of the forest for $R_0 = 3$~Mpc from the emission redshift, as
$\pm 1\sigma$ envelopes around the expected $\tau_\mathrm{eff}$ in the case of
no overdensity ($\Xi =1$). }
\label{fig:overdensity}
\end{figure*}

\item From the corrected $\tau^*(z)$ we computed the effective optical depth
in bins along the line of sight and divided it by the expected values at the
appropriate values of $z$ (given by eq.~\ref{eq:tau}).  This procedure had to
be carried out for each object separately.  We define this ratio of effective
optical depths as the \emph{overdensity degree} $\Xi$. We performed this
calculation for different bin sizes, corresponding to increments of 2, 3, 5
and 8 proper Mpc. Qualitatively, the results are the same, with
a bin size of 3~Mpc showing the clearest trends. We only quote results for
that bin size in the following.

\item We then merged all lines of sight, resulting in 40 estimates of $\Xi$
per distance bin. In each bin we estimated the mean and the standard
deviation of $\Xi$. We used our Monte-Carlo simulated spectra to reproduce 
our QSO sample and thereby estimated the amount of variance in $\Xi$
attributed to pure Poisson noise. 
\end{enumerate}

\subsection{Overdensity distribution}\label{txt:ovdensdist}

The dependence of the average $\Xi$ as a function of distance is presented in
Fig.~\ref{fig:xifluctuation} (solid line).  At large distances ($>5$~Mpc),
$\Xi$ fluctuates around unity as predicted by perfect Poissonian noise. Only
in the innermost bin for $\la 3$~Mpc does the mean value of $\Xi$ increase 
towards an overdensity degree of a factor of 2 (in units of optical depth).
So far this can be seen as supporting the notion of quasars \emph{typically}
residing in overdense regions \citep[e.g.][]{dodorico08}. However, in the same panel we also show the
average $\Xi$ if the 10 objects (25~\% of the sample) with the highest
individual overdensities are removed (see below for details of this
deselection).  For all bins beyond $\sim 5$~Mpc, the run of $\Xi$ with
distance is statistically indistinguishable for the two samples. Only for the
nearest two bins the two profiles deviate, in the sense that the dotted
profile (based on 75~\% of the sample) is fully consistent with no
overdensities at all.

Figure~\ref{fig:overdensity} shows the individually measured `overdensity
degrees' within the innermost 3~Mpc bin as a function of quasar luminosity at
the Lyman limit, quasar redshift, and proximity effect strength $\log a$ (cf.\
Sect.~\ref{txt:pepd}), respectively.  This diagram reveals that the corrected effective
optical depths for individual quasars range from close to zero to up to 6 times
larger than in the mean Ly$\alpha$ forest.  The horizontal dashed lines again
indicate the amount of scatter ($\pm 1\sigma$) expected for pure Poissonian
variance within the same physical scale around each of the QSOs.

Panel [a] shows that there is no systematic tendency of $\Xi$ to vary with the
quasar Lyman limit luminosity. The spread appears to be somewhat larger at
high luminosities, which may reflect the expected trend of luminous quasars to
reside in more massive environments. But there are also several high-$L$
objects with $\Xi$ values around unity.

Regarding the redshift dependence depicted in Panel [b] of
Fig.~\ref{fig:overdensity}, no significant  trend can be seen here either.
There may be a deficit of high overdensities for $z > 3$, but this could
equally be an effect of small number statistics.

However, panel [c] of Fig.~\ref{fig:overdensity} reveals some dependence
between $\Xi$ and proximity effect strength $\log(a)$. The four QSOs with 
$\Xi > 3$ are among the six objects with the weakest proximity effect (highest 
values of $\log a$). A quantitative test using the standard linear correlation 
coefficient $r$ shows that $\log\Xi$ is indeed correlated with $\log a$, $r=0.6$ (with a
probability of $< 1$~\% of no correlation), but that this trend is strongly driven by the 
few points with $\Xi > 3$. Without these points, the correlation coefficient reduces to 
$r = 0.4$, and there is a 3~\% probability of no intrinsic correlation.

\subsection{Effects of overdensities on the proximity effect strength distribution}\label{txt:ll_z_dep_overd}

We now investigate how the presence of some fraction of the QSOs residing in
dense environments changes the properties of the PESD.  We generated a new
sample of synthetic spectra at $z=$ 2, 3 and 4, respectively, at a fixed
luminosity of $\log(L_{\nu_0}/\mathrm{erg\ s^{-1}\ Hz^{-1}})=31.5$, with the
difference that we now required that our sight lines show excess absorption
near the emission redshift. We did not assume any particular radial
distribution for the overdensity profile, but simply continued to populate the
spectrum within 3~Mpc with lines until the effective optical depth reached a
given threshold. We constructed two samples, for overdensities of $\Xi=2$ and
4, respectively, to approximately mimic our observational results (500 spectra
for every $\Xi$).

\begin{figure}
\includegraphics[width=\hsize]{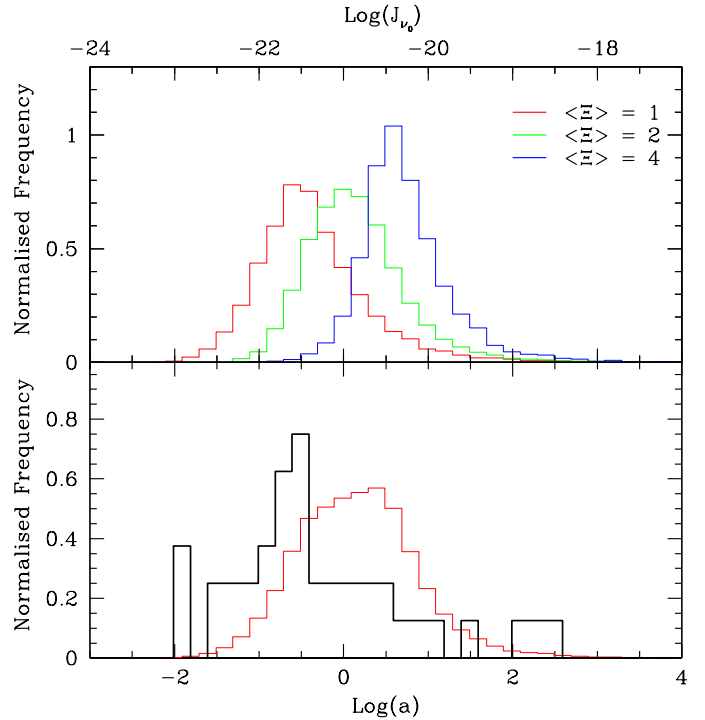}
\caption{Top panel: Expected PESD computed from Monte-Carlo simulations 
for different expectation values of the overdensity degree ($\Xi = 1$, 2, 4,
respectively). The main effect is a substantial shift towards an on average
weaker proximity effect. Bottom panel: Comparison of the observed PESD and the
expectation for an assumed Gaussian distribution of overdensities (see text
for details). 
}
\label{fig:pesd_overd}
\end{figure}

These simulations show that if all quasars had the same enhanced expectation value $\left<\Xi\right>$, the
\emph{shape} of the PESD would not be greatly affected for all studied redshifts.
The main effect would be a global shift of the PESD towards higher $\log a$ values,
with the shift increasing as $\left<\Xi\right>$ becomes larger. This is shown in the top
panel of Fig.~\ref{fig:pesd_overd}, where we plotted the PESD for three
expectation values of $\left<\Xi\right> = 1$, 2, and 4, respectively. 
Already for an average enhancement of $\left<\Xi\right> = 2$, the peak of the
PESD is shifted by $\sim 0.6$~dex.

It follows that if quasar environments are systematically overdense by some
uniform factor, we would be unable to tell this from our data alone: Our empirical
determination of $\Xi$ depends on assuming the correct global $J_{\nu_0}$.  If
the mode of the PESD is biased by some factor, we would \emph{underestimate}
the $\Xi$ values of the individual quasars by the same factor, effectively
renormalising the \emph{observed} $\Xi$ scale to an expectation value of
unity. Of course, this is nothing but yet another manifestation of the
well-known degeneracy between the UVB and overdensities around quasars.

However, such a simple picture of all quasars having the same density
enhancement factor would be highly unrealistic. If quasars reside in
overdensities, the enhancement will have a distribution stretching over at
least a factor of few.  Because the shift of the PESD depends on
$\left<\Xi\right>$, this will lead to a broadening of the PESD by an amount
depending on the physical spread of overdensities. Notice that this spread
will come on top of the inevitable scatter due to Poissonian variance. As an
example toy model, we show in the bottom panel of Fig.~\ref{fig:pesd_overd}
the modified PESD for an assumed physical rms scatter in $\left<\Xi\right>$ of
0.25~dex. The predicted distribution is considerably different; it is much
broader and more box-shaped, and certainly much less in agreement with the
observed PESD than the one for $\left<\Xi\right> = 1$. While the assumption of
a scatter of 0.25~dex is entirely ad-hoc, the example shows that the PESD
reacts quite sensitively, and a very small physical dispersion around the
expectation value of $\Xi$ is required to make observed and predicted
PESD consistent. This is naturally obtained only if the expectation value for
most quasars is close to $\left<\Xi\right> \approx 1$.

These considerations strongly support the notion that there is no strong
\emph{systematic} overdensity bias for the proximity effect. While the
excellent agreement between our modal estimate and other ways of determining the
UVB intensity (Fig.~\ref{fig:uvb_evolmode}) is already highly suggestive, we
cannot of course use this agreement both to justify our method \emph{and}
regard it as an indepenent measurement of the UVB.  The comparison of expected
and observed PESD shapes adds independent information from the distribution
properties and breaks the UVB-overdensity degeneracy.

Clearly some fraction of objects sits in overdense regions. This, however, only
produces a tail to the PESD by adding some high values of $\log a$.  A
comparison of Fig.~\ref{fig:overdensity}c with Fig.~\ref{fig:log_a_data} shows
that the objects with strong overdensities are exactly those lying outside the
prediction for the simple Poissonian model.

We note in passing that there is no reason to interpret the few quasars
showing an extremely strong proximity effect as sitting in
`underdensities'. The three objects with the lowest $\log a$ values have a
mean $\Xi$ very close to unity (see Fig.~\ref{fig:overdensity}c), and only one 
object is located outside of the $\pm 2\sigma$ envelope of $\Xi$ expected
for pure Poissonian fluctuations around the mean density.

\subsection{A hybrid method of estimating the UV background} \label{txt:hybrid}

In the following we adopt the notion that only a relatively small fraction of
at most $\sim 25$~\% of our quasars is affected by significant overdensities.
If this is true, then for a completely unbiased estimation of the UVB we need to: 
(i) remove the quasars with biased environments;
(ii) correct for the asymmetry bias in the proximity effect strength
distribution. 

In order to conduct the first step we had to identify outliers in the $\Xi$
distribution. However, recall that $\Xi$ is based on the assumption that we already 
know the UVB background. As an initial guess we adopted the modal value
from the PESD, from which we then could compute $\Xi$ for all quasars.  There is
no obvious cutoff, and many quasars have $\Xi$ values that are consistent with
weak overdensities. We decided to make a conservative cut in the sense
that we would rather avoid too many even mildly overdense objects. We set the
threshold at $+1\sigma$ (predicted scatter for Poissonian variance), implying that
quasars with $\Xi > \Xi_0=1.8$ would be excluded from the analysis. This
was the case for 10 objects, or 25~\% of the sample. 

We then constructed the PESD based on the remaining 30 objects. The modal
value remains unchanged, but the uncertainty obviously increases 
(0.2 rather than 0.15 dex). The fact that the mode of the PESD is robust
against the inclusion or exclusion of 25~\% of the sample is certainly
encouraging. It is also clear that it would not be possible to use a much smaller fraction of objects, 
as the uncertainties of estimating the mode would rapidly go up.

Next, we determined the systematic offset of the `combined sample averaging'
estimator (used in Sect.~\ref{txt:comb_PE}) due to the asymmetry of the PESD. We proceeded as follows:
We generated mock quasar samples from the Monte Carlo simulated spectra,
assuming no systematic overdensity bias, with the same redshift and luminosity
distributions as the observed sample. We removed those objects showing
apparent overdensities (occurring because of statistical noise) and determined
the best-fit UVB value using the standard sample combination method as
described in Sect.~\ref{txt:comb_PE}.
Repeating this step 500 times, we averaged 
over the differences to the input UVB, obtaining a mean
offset of $\Delta \log a = 0.27$~dex.

Finally, we conducted the same procedure for the observed sample 
with $\Xi<1.8$, performing a least-squares fit of Eq.~\ref{eq:fit} 
to the data and correcting the resulting value by the above offset. 
The inferred value is $\log J_{\nu_0} = -21.46^{+0.14}_{-0.21}$, which
we consider as the best and most robust estimate of the global UV background
intensity for our sample.

In order to find out how much the method depends on the
adopted criterion to eliminiate outliers,
we repeated the described procedure once more, but now setting the
threshold at $+2\sigma$, corresponding to $\Xi_0=3.1$. This implies
that only the four objects with the highest $\Xi$ values were excluded
from our sample. We redetermined the mean offset, which remained
unchanged. The least-squares fit of Eq.~\ref{eq:fit} to the data, corrected 
for the relative offset, yields again a value of $\log J_{\nu_0} = -21.49^{+0.14}_{-0.21}$,
demonstrating that our hybrid method is highly robust and quite
insensitive to the outlier criterion. 

Determining the UVB intensity using this hybrid method has one important advantage over 
the modal PESD estimator: It depends less critically on the sample
size. The mode is a robust statistical quantity only if the histogram from which it is
estimated is reasonably well defined. In the next section we will split the 
sample into subsets to investigate the redshift evolution of the UVB. It will
be seen that the PESD of such subsamples can become rather broad, inhibiting
any confident mode estimation. The hybrid technique, on the other hand,
involves straightforward averaging and therefore also works for smaller samples.

Table~\ref{tab:uvb_res} summarises all our different methods of estimating the 
UVB intensity and the resulting values.

\begin{table}
\centering
\caption{Summary of our UV background measurements and employed methods, 
for the sample median redshift $z=2.73^\dagger$.}
\label{tab:uvb_res}
\begin{tabular}{lc}
\hline\hline\noalign{\smallskip}
Method& $\log J_{\nu_0}$ \\
\noalign{\smallskip}\hline\noalign{\smallskip}
fit to combined sample   & $-21.10^{+0.14}_{-0.22}$ \\
\textbf{mode of PESD}               & $\mathbf{-21.51\pm0.15}$ \\
mean of individual $\log a$ & $-21.24\pm0.17$ \\
fit to combined subsample ($\Xi<1.8$) & $-21.19^{+0.14}_{-0.21}$ \\
\textbf{fit to combined subsample ($\mathbf{\Xi<1.8}$,corrected)}   & $\mathbf{-21.46^{+0.14}_{-0.21}}$ \\
\noalign{\smallskip}\hline\noalign{\smallskip} 
\end{tabular}
\begin{list}{}{}
\item[$\dagger$:] The two boldfaced rows represent the unbiased estimates.
\end{list}
\end{table}

\section{The redshift evolution of the UV background}\label{txt:uvb}

The high S/N of our data and the size and redshift range of the sample
prompted us to explore whether we could split the sample into subsets of
different redshifts and still get meaningful constraints on the UVB.

\begin{figure}
\resizebox{\hsize}{!}{\includegraphics{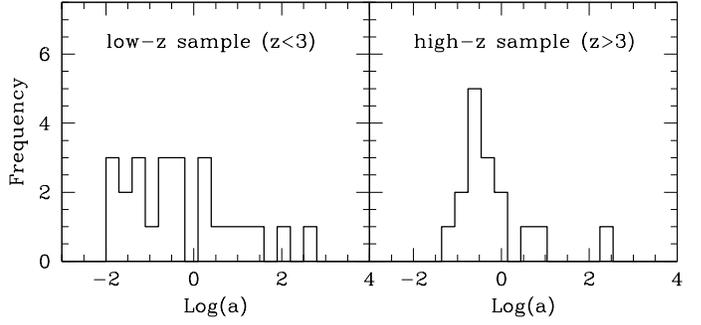}}
\caption{Observed PESD for our sample of quasars divided into two redshift intervals. The shapes of the 
two PESDs differ substantially from each other as predicted in Sect.~\ref{txt:ll_z_dep}.}
\label{fig:logahilo}
\end{figure}

As a first attempt we divided the sample into two subsets with $z<3$ and
$z>3$. For each subset we constructed the PESD separately, shown
in Fig.~\ref{fig:logahilo}. The distributions look substantially different
from each other. The PESD of the lower redshift subset is very broad, and no
clear peak can be recognised. On the other hand, the high redshift PESD is
significantly narrower and displays a well-defined maximum. The overall
behaviour is qualitatively consistent with the predictions of
Fig.~\ref{fig:log_a_distMC} based on our Monte Carlo simulations, where we
found that the widths of the PESD should decrease as redshift increases
(Sect.~\ref{txt:ll_z_dep}).  The modal value obtained from the high redshift
sample is $\log J_{\nu_0} \simeq -21.6$, slightly lower than the overall best
estimate.  In the lower redshift subset, even a rough guess of the
distribution mode is impossible.

\begin{table}
\small\centering
\caption{UV background intensity as function of redshift for the set of quasars that show 
overdensity degree $\Xi \le 1.8$ as plotted in Fig.~\ref{fig:uvb_evol}.}
\label{tab:uvb_evol}
\begin{tabular}{ccc}
\hline\hline\noalign{\smallskip}
$<z>$&  $\log J_{\nu_0}$ & $\sigma_{\log J_{\nu_0}}$\\
\noalign{\smallskip}\hline\noalign{\smallskip}
 2.10   & -21.15  &  0.20 \\
 2.30   & -21.89  &  0.20 \\
 2.50   & -21.20  &  0.19 \\
 2.70   & -21.62  &  0.19 \\
 2.90   & -21.46  &  0.20 \\
 3.10   & -21.50  &  0.20 \\
 3.30   & -21.91  &  0.20 \\
 3.55   & -21.61  &  0.20 \\
 3.80   & -21.71  &  0.21 \\
\noalign{\smallskip}\hline\noalign{\smallskip} 
\end{tabular}
\end{table}

\begin{figure*}
\sidecaption
\includegraphics[width=12cm]{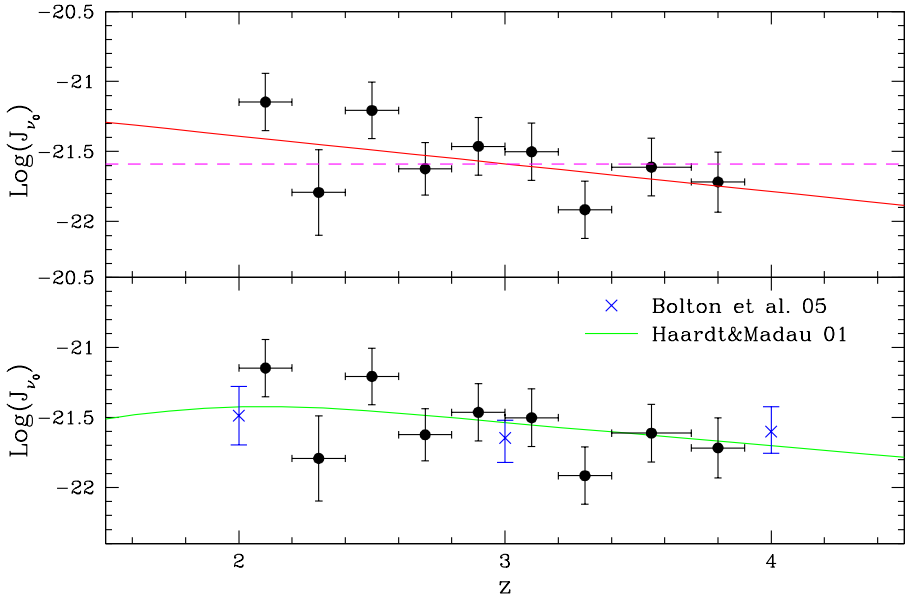}
\caption{Evolution of the mean intensity of the UV background with redshift from those quasars with 
overdensity degree $\Xi<1.8$ (see text for details). We estimated the UVB using the combined analysis 
of the proximity effect in $\Delta z=0.2$ intervals over the range $2 \lesssim z \lesssim 4$. 
\emph{Top panel}:  $\log J_{\nu_0}(z)$ in comparison to a constant evolutionary model (dashed) 
and a linear fit to the data (solid).  \emph{Bottom panel}:  $\log J_{\nu_0}(z)$ in comparison 
to the model by \citet{haardt01} and the more recent determination of the UVB evolution from 
\citet{bolton05}. The errors in the $z$ direction represent the range used to measure the average UVB.}
\label{fig:uvb_evol}
\end{figure*}

A more powerful way of estimating the UVB intensity is the hybrid method
described in the previous subsection. We now employed this method to
investigate the redshift evolution of the UVB. After merging all quasar 
spectra in the same way as for measuring the evolution of the Ly$\alpha$ 
effective optical depth (Sect.~\ref{txt:tau_evol}) we sampled the 
Ly$\alpha$ forest in redshift bins of $\Delta z = 0.2$. We used only 
the forest in the redshift range $2 \la z \la 4$ in order not to be 
affected by small number statistics. For every bin we fitted Eq.~\ref{eq:fit} 
to the data and obtained a preliminary estimate of $J_{\nu_0}$. 
These values are still systematically high, both as a result of the PESD asymmetry 
and because of the influence of overdense lines of sight. In particular, 
the redshift ranges at $z \sim 2.2$ and at $z\sim 2.9$ are heavily affected
by the latter effect, as at these redshifts we find the majority of objects 
with a strongly overdense environment (cf.\ Fig.~\ref{fig:overdensity}). 
In such narrow redshift bins, the values of the UVB are extremely sensitive 
to locally enhanced absorption, causing deviations of up to 1.3 dex. 

We now followed the same recipe as in Sect.~\ref{txt:hybrid} above: We removed
from the sample those objects with $\Xi > 1.8$ within the innermost 3~Mpc. We
then applied the combined average estimator for each redshift interval and
applied the PESD asymmetry bias determined from the simulations. We obtained
in total eight values of $J_{\nu_0}$, for eight different redshift bins as
listed in Table~\ref{tab:uvb_evol}. These result are presented as black points in
Fig.~\ref{fig:uvb_evol}, where the horizontal bars indicate the redshift
interval and the vertical bars are the statistical errors.

The top panel of Fig.~\ref{fig:uvb_evol} shows that our measurements do
suggest a certain evolution of the UVB intensity with redshift. Fitting 
a simple least-squares linear relation to the data points we get
\begin{equation}
\log J_{\nu_0}(z) = (-0.20\pm 0.14)\ z + (-21.0\pm0.4)\:.
\end{equation}
This fit (depicted as the solid line in the top panel of
Fig.~\ref{fig:uvb_evol}) yields residuals fully consistent with random
errors. However, the slope is not very steep, and in view of the size of the
error bars it is appropriate to ask whether evolution is actually
\emph{demanded} by the data. For comparison we considered also the null
hypothesis of no evolution, which is represented by the horizontal dashed line
in Fig.~\ref{fig:uvb_evol}). The quality of that model is somewhat poorer, but
a KS test gives still 10~\% acceptance probability, which is at best a
marginal rejection. We can therefore only tentatively claim to detect the UVB
evolution within this redshift interval.  In a
different take, we can confidently exclude a very steep redshift evolution,
such as would be predicted from a UV background made predominantly of quasar
light.

How do these results compare with other works aimed at estimating or
predicting the UVB evolution at similar redshifts? We first consider the
prediction by \citet{haardt01}. They constrained $J_{\nu_0}(z)$ by integrating
the contributions of the observed quasar and young star-forming galaxy
populations, accounting also for intergalactic absorption and re-emission.
Their predicted redshift evolution of the UVB intensity is depicted as the
solid green line in the lower panel of Fig.~\ref{fig:uvb_evol}). The agreement
with our data points could hardly be better; in particular, the shallow slope
is almost exactly reproduced. We roughly estimate a slope of
$\mathrm{d}J/\mathrm{d}z = -0.16$ for the Haardt \& Madau model at $z \ga
2.2$, which is extremely close to our best-fit value of $-0.2$. 
Nevertheless it should be emphasised that the galaxy contribution to the 
UVB crucially depends on the poorly constrained escape fraction of Lyman continuum 
photons from galaxies \citep[e.g.][]{shapley06,gnedin07}.

Recently, several studies calibrated the outcome of hydrodynamical IGM
simulations through the measured Ly$\alpha$ forest opacity to estimate the
photoionisation history of intergalactic hydrogen
\citep[e.g.,][]{bolton05,giguere08b}. These studies typically gave an
approximatly constant UVB intensity within the range $2<z<4$.
As an example we show the results of \citet{bolton05} by the crosses 
in the lower panel of Fig.~\ref{fig:uvb_evol}). 
As stated above, a no-evolution scenario as found by these studies 
is also consistent with the errors, and our results are also consistent with
those of \citet{bolton05} and \citet{giguere08b}. 

At any rate, our measurements confirm the prevailing assumption that
quasars alone are unable to keep the IGM at a highly ionised state at
redshifts larger than three. The integrated UV emissivity of AGN 
has a maximum very close to $z=2$ and drops by about an order of magnitude
towards $z=4$ \citep{wolf03,hopkins07}, whereas we find the UVB intensity
to fall by only $\sim 0.2$~dex (2$\sigma$ limit is 0.48~dex). 
This strengthens the notion for a non-negligible contribution by 
star-forming galaxies to the UV background radiation field; their
fractional contribution increases when going to higher redshifts. 

We can make a simple back-of-an-envelope estimate of the mixing ratio
between quasars and stars by assuming that the absolute UV emissivity of
star-forming galaxies is roughly constant with redshift between $z=2$ and
$z=4$. Since the integrated quasar emissivity at $z=4$ falls to $\sim 10$~\%
of its $z=2$ value \citep{hopkins07}, and using our best-fit measurement of
the overall UV background evolution (adopting a factor 0.4 between $z=2$ and
$z=4$), we obtain that the H$^0$ photoionisation rate at $z=2$ is dominated by
quasars by a factor of $\sim 2$ over star forming galaxies. At $z=4$,
galaxies would then dominate by a factor of 5 over quasars. Any change towards
making the total UVB evolution more constant with redshift would increase the
fraction of starlight in the UVB.

\section{Conclusions}\label{txt:concl}

We have analysed the largest sample of high-resolution quasar spectra
presented in a single study to date. We demonstrated that the line-of-sight
proximity effect is a universal phenomenon that can be found in the spectrum
of essentially every individual high-redshift quasar. While in Paper~I we arrived
at the same conclusion using low-resolution spectra, we now can significantly
reinforce this claim. In particular, continuum placing uncertainties, which
were a bit of an issue in Paper~I, play absolutely no role now, given the
high spectral resolution and high S/N data exploited in this paper.

Our study rehabilitates the proximity effect as a tool to investigate the
intensity at 1~Ryd (more accurately: the H$^0$ photoionisation rate) of the
cosmic ultraviolet background. Previous investigations using the proximity
effect nearly always produced UVB estimates that were suspiciously high and
indeed at variance with other means to estimate or predict the UVB.  It has
been suggested in the literature that quasars typically reside in denser than
average large-scale environments, and that therefore excess absorption biases
the proximity effect to appear too weak, resulting in an overestimated UVB.
We argue that this is not the main reason for the discrepancies
between the proximity effect and other methods.

The actually measured strength of the proximity effect in a single quasar line
of sight depends on the UVB, but also on the number of absorption lines
present in the `proximity effect zone'. That number is generally small, and to
first approximation the presence or absence of a line at a given redshift can
be described as a Poissonian random process. We have demonstrated that the
distribution of the resulting `proximity effect strength' parameter (which
directly relates to the UVB) is significantly skewed, even without invoking
any physical overdensities. Any direct averaging over a sample of sightlines
will inevitably bias the resulting UVB. Only by looking at individual lines of
sight separately can this effect be detected and removed.

We proposed and used two different methods of estimating the UVB intensity in
the presence of this bias. In the first method, the UVB is taken directly from
the mode of the proximity effect strength distribution (PESD). This method is
simple and robust against a small number of outliers; on the other hand it is
limited to large samples and depends on the condition that the mode is not
affected by the outliers. Our second method involves Monte Carlo simulations
to calibrate the asymmetry bias of the PESD and then uses straightforward
sample averaging, possibly combined with a prior rejection of outliers, and
the result is then corrected for the bias. This `hybrid' method is quite
stable and works also for smaller samples; it is, however, more complicated to
use and depends on a successful removal of outliers.

Our best estimates of the UVB intensity using these two methods give very
similar results. The modal estimate is $\log J_{\nu_0} = -21.51\pm 0.15$,
and the hybrid estimator gives $\log J_{\nu_0} = -21.46^{+0.14}_{-0.21}$,
which is statistically indistinguishable. This value is in excellent
agreement with other methods to estimate or predict the UVB.

If the UV background is known, the data used for the proximity effect can also
be employed to reconstruct the absorption pattern without the radiative
influence of the quasar. Doing this for our sample of 40 quasars we found that
only 10~\% of them showed very significant excess absorption (by more than a
factor of $\sim 3$) while most are distributed in a way more or less
consistent with Poissonian fluctuations around the cosmic mean absorption
expected at these redshifts. Given the widespread notion of quasars residing in
massive galaxies, which in turn should sit in massive haloes with
significantly clustered environments, this sounds surprising. We have no
intention of questioning that picture, and we stress that our measurements
constrain exclusively the neutral hydrogen distribution averaged on scales of
$\sim 3$~proper Mpc towards the quasar. 

Nevertheless, our results may bear some relevance to the question of what are the
typical halo masses for quasars. Most of the QSOs in our sample are
optically selected and therefore presumably radio-quiet, which could mean that
their haloes are not exceptionally massive. In a recent paper,
\citet{mandelbaum08} estimated halo masses of AGN in the Sloan survey from
clustering and galaxy-galaxy lensing, finding that optically selected AGN
follow the same relation between stellar and halo mass as normal galaxies, 
whereas radio-loud AGN have much higher halo masses. (Incidentally, the
outstanding object in our sample in terms of overdensity is the radio quasar
PKS~0237$-$233; but the next-ranked objects are all optically selected.)

Our argument against significant overdensities of quasars and in favour of the
applicability of the proximity effect to measure the UV background gains
weight by the excellent agreement between our estimates of the UVB intensity
and those obtained by completely different methods. This notion is supported
independently by considering the \emph{shape} of the PESD. The observed
distribution width clearly favours average densities close to unity; if
significant overdensities were involved, a much broader distribution would be expected than 
the one observed. This additional constraint breaks the degeneracy
between UV background and overdensities that has haunted the field for so
long.

Finally, we also attempted to constrain the redshift evolution of the
UVB intensity. Although the results are still quite uncertain, we could for
the first time from a single quasar sample derive useful limits on the amount
of evolution. We can rule out at high significance that quasar light dominates
the UVB at redshifts $z\ga 4$. Our best fit suggests a mild decrease in the
UVB intensity towards higher redshifts, and the derived slope is in
astonishingly good agreement with the predictions by
\citet{haardt01}. Nevertheless, an approximately constant UVB as found by
\citet{bolton05} and others is also formally consistent with our data.

We have shown that the proximity effect holds the potential to derive
important cosmological quantities. While our sample is large in comparison
to those of previous studies using high-resolution spectra, it still
suffers in several aspects from small number statistics. As the public
archives are growing, substantial progress can be expected from applying 
the proximity effect analysis to further spectra, in particular at higher
redshift. 

\begin{acknowledgements}
We would like to thank ESO for making the ESO data archive publicly available. We also thank
Tae$-$Sun Kim for assisting us in the data reduction process.
A.D. and G.W. were partly supported by a HWP grant from the International Helmholtz Institute for 
Supercomputational Physics. We finally acknowledge support by the Deutsche Forschungsgemeinschaft 
under Wi~1369/21-1.
\end{acknowledgements}

\bibliography{bibliography}

\Online\onecolumn

\begin{appendix}

\section{Proximity effect on individual lines of sight}\label{app:PElos}

\begin{figure*}\centering
\includegraphics*[width=17.2cm]{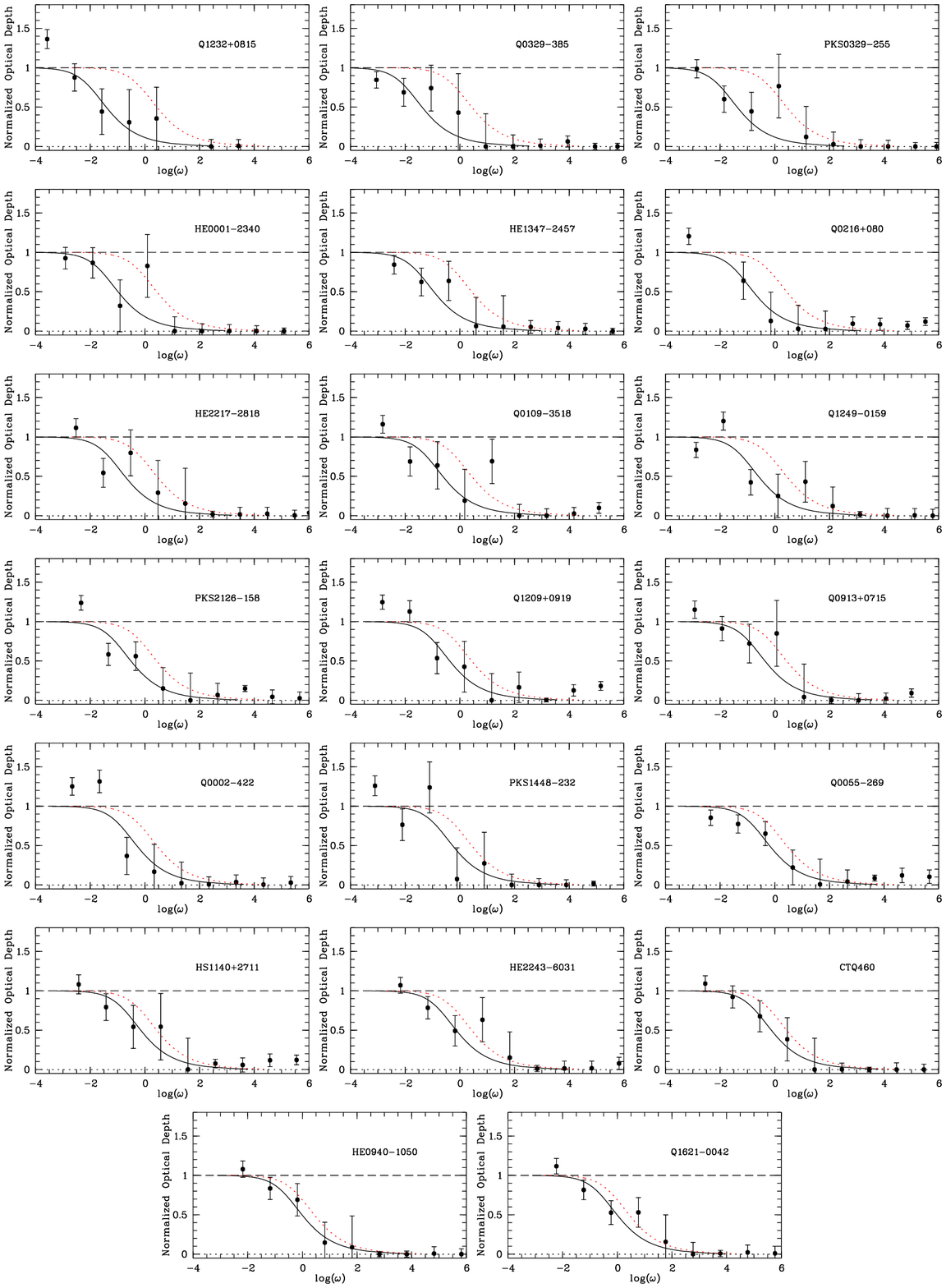}
\caption{The proximity effect signatures in individual lines of sight. Each panel shows the normalised optical depth $\xi$ versus $\omega$, binned in steps of $\Delta\log\omega = 1$, with the best-fit model of the combined analysis superimposed as dotted red lines. The solid lines delineate the best fit to each individual QSO as described in the text.}
\label{fig:PE_prof_sample1}
\end{figure*}

\addtocounter{figure}{-1}
\begin{figure*}\centering
\includegraphics*[width=17.2cm]{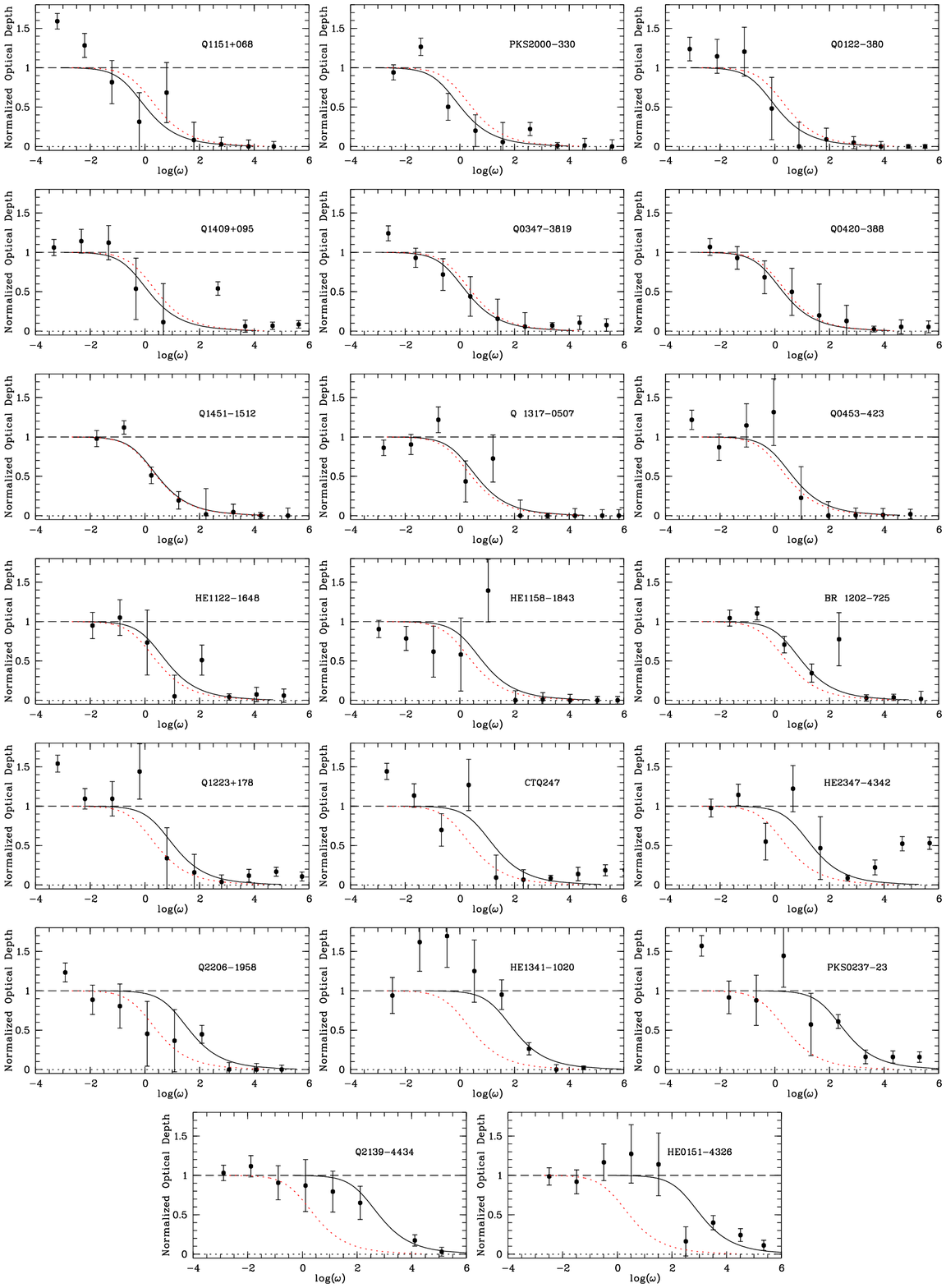}
\caption{Continued}
\end{figure*}
\end{appendix}

\end{document}